\documentclass[conference]{IEEEtran}
\IEEEoverridecommandlockouts
\usepackage{cite}
\usepackage{amsmath,amssymb,amsfonts}
\usepackage{algorithmic}
\usepackage{graphicx}
\usepackage{textcomp}
\usepackage{xcolor}

\usepackage{paralist}
\usepackage{enumitem}
\usepackage{graphics}
\usepackage{dblfloatfix}
\usepackage{url}
\usepackage{tabularx}
\usepackage{caption}
\usepackage{subcaption}
\usepackage{xspace}
\usepackage[colorinlistoftodos,prependcaption,textsize=tiny]{todonotes}
\usepackage{hyperref}
\usepackage[ruled, vlined]{algorithm2e}

\def\BibTeX{{\rm B\kern-.05em{\sc i\kern-.025em b}\kern-.08em
    T\kern-.1667em\lower.7ex\hbox{E}\kern-.125emX}}

\newcommand\inlinecomment[1]{ }

\newcommand{\sws}{{shallow water simulation}\xspace}
\newcommand{\swe}{shallow-water equations\xspace}
\newcommand{\blaz}{Blaz\xspace}
\newcommand{\pyblaz}{\mbox{PyBlaz}\xspace}

\newcommand{\nan}{\texttt{NaN}\xspace}

\newcommand{\half}{\texttt{FP16}\xspace}
\newcommand{\single}{\texttt{FP32}\xspace}
\newcommand{\double}{\texttt{FP64}\xspace}
\newcommand{\bfloat}{\texttt{bfloat16}\xspace}

\newcommand{\byte}{\texttt{int8}\xspace}
\newcommand{\short}{\texttt{int16}\xspace}
\newcommand{\inttt}{\texttt{int32}\xspace}  % int thirty-two
\newcommand{\intsf}{\texttt{int64}\xspace}  % int sixty-four

\begin{document}

\title{What Operations can be Performed Directly on Compressed Arrays, and with What Error?\thanks{An extended but earlier version of paper in \url{https://dl.acm.org/doi/10.1145/3624062.3625122}, published at the DRBSD Workshop}\\
% {\footnotesize \textsuperscript{*}Note: Sub-titles are not captured in Xplore and
% should not be used}
% \thanks{Identify applicable funding agency here. If none, delete this.}
}
%~~~

%~~~

\author{\IEEEauthorblockN{Tripti Agarwal$^{*}$}
\IEEEauthorblockA{\textit{Kahlert School of Computing} \\
\textit{University of Utah}\\
Salt Lake City, USA \\
tripti.agarwal@utah.edu}
\and
\IEEEauthorblockN{Harvey Dam$^{*}$\thanks{* Equal contribution with Agarwal}}
\IEEEauthorblockA{\textit{Kahlert School of Computing} \\
\textit{University of Utah}\\
Salt Lake City, USA \\
harvey.dam@utah.edu}
\and
\IEEEauthorblockN{Dorra Ben Khalifa}
\IEEEauthorblockA{\textit{dept. name of organization (of Aff.)} \\
\textit{University of Perpignan}\\
Perpignan, France \\
dorra.ben-khalifa@univ-perp.fr}
\and
\IEEEauthorblockN{Matthieu Martel}
\IEEEauthorblockA{\textit{dept. name of organization (of Aff.)} \\
\textit{University of Perpignan}\\
Perpignan, France \\
matthieu.martel@univ-perp.fr}
\and
\IEEEauthorblockN{P. Sadayappan}
\IEEEauthorblockA{\textit{Kahlert School of Computing} \\
\textit{University of Utah}\\
Salt Lake City, USA \\
saday@utah.edu}
\and
\IEEEauthorblockN{Ganesh Gopalakrishnan}
\IEEEauthorblockA{\textit{Kahlert School of Computing} \\
\textit{University of Utah}\\
Salt Lake City, USA \\
ganesh@utah.edu}
}

\maketitle

\begin{abstract}
In response to the rapidly escalating costs of computing with large matrices and tensors caused by data movement, several lossy compression methods have been developed to significantly reduce data volumes. Unfortunately, all these methods require the data to be decompressed before further computations are done. In this work, we develop a lossy compressor that allows a dozen fairly fundamental operations directly on compressed data while offering good compression ratios and modest errors. We implement a new compressor \pyblaz based on the familiar GPU-powered PyTorch framework, and evaluate it on three non-trivial applications, choosing different number systems for internal representation. Our results demonstrate that the compressed-domain operations achieve good scalability with problem sizes while incurring errors well within acceptable limits. To our best knowledge, this is the {\em first such lossy compressor} that supports compressed-domain operations while achieving acceptable performance as well as error.
%
% As the amount of data is increasing it is important to compress the data and use compressed-space operations to reduce the expense of decompression. The aim of this research is to investigate the impact of using compressed-space operations on floating-point arrays. We create a modified version of the already available floating-point compressor and add various compress space operations. We then test our work on varied datasets to see how these compressed-space operations can be used to gain insights/ features from the dataset. We also test our compressed-space operations with uncompressed data to show how much error gets accumulated due to compression and the operations we perform on top of it. 
%
\end{abstract}


\section{Introduction}

% {\bf GG and MM: Motivate array programming as a part of PL.}
% Intro tone for the array paper, given that it is a workshop paper where "array programming" has to be motivated to a crowd of PL-ish people. 

% From their web overview

% \begin{quote}\scriptsize
%     The ARRAY series of workshops explores all aspects of array programming, such as languages, formal semantics, array theories, productivity/performance tradeoffs, libraries, notation such as including axis- and index-based approaches, intermediate languages, and efficient compilation.

%     IN THE CFP, THESE BULLETS ALSO ARE STATED AND WE CAN WRITE TOWARD THEM

    % representation of and automated reasoning about mathematical structure, such as static and dynamic sparsity, low-rank patterns, and hierarchies of these, with connections to applications such as graph processing, HPC, tensor computation and deep learning;

    % ALSO

    % efficient mapping of array programs, through compilers, libraries, and code generators, onto execution platforms, targeting multi-cores, SIMD devices, GPUs, distributed systems, and FPGA hardware, by fully automatic and user-assisted means.

    % AIM FOR a 7-page paper w/o refs

    % FIGURES are important 
    
% \end{quote}

 % HPC and data sets are not comparable

Today's computational
processes in areas such as
high-performance computing (HPC)
and machine learning (ML) consume as well as produce large multi-dimensional matrices and tensors,
making it imperative that we rein in the volume of data moved.\footnote{Data locality optimization also helps reduce data movement through better reuse; this is not considered in our work.}
It has been demonstrated that 
  lossy compression methods (elaborated later in Section~\ref{sec:background}), can achieve significant data volume reduction---anywhere from a factor of 8 to 60 or more.
These methods have been used to reduce the volume of data checkpointed and even to avoid  recomputing frequently reused data: this data can be stored in compressed form, and the compression costs are amortized over multiple reuses.
While
sufficient details of the original data 
 are available when the compressed data
 is decompressed, 
 {\em existing methods require that the
 data be decompressed before operating}.
In this work, we show that through a suitable modification of existing data compression pipelines,   a dozen 
commonly needed operations can be performed directly without decompression---with many operations incurring no additional error than sustained during compression.

While such direct operation has been attempted in limited settings before
(Section~\ref{sec:background}) and is also found in methods such as  
homomorphic encryption~\cite{homomorphic-encryption},
ours is the first proposal that significantly advances the available repertoire of operations while also (1)~providing an easily accessible and reasonably efficient implementation on top of GPU-powered PyTorch, and (2)~studying the performance and error of direct operations in a variety of nuanced ways.

To provide an adequate context for detailing our work, let us consider
an obvious alternative:
 precision tuning where
  one chooses \half or \single (IEEE floating-point standard 16-bit number or 32-bit number~\cite{ieee-754-1985}) in lieu of 
\single or \double, respectively.
While almost a  drop-in replacement (modulo recompilation, etc.) to an existing piece of code,
precision tuning  
attains a compression ratio
of only two to four.
Unfortunately, one can almost never
be sure which few words are about to
overflow or underflow: all other words may still have unused dynamic range.
This is one reason why, despite all the hoopla, precision tuning has seen scanty update in HPC.\footnote{The use in ML falls into the territory of controlling the number of NaNs, loss-scaling, etc.~\cite{apex}---beyond what we consider here.}

Lossy data compression operates by taking a data volume and distributing the error.
As one example, some methods take the largest magnitude exponent in the volume, shift the remaining values right (thus losing some low-significance bits in the process) to match this exponent, and then {\em encode} the remaining data using an orthonormal transform.
Thereafter, only the {\em compression coefficients} are kept---not the original data.
Lossy compression is not a drop-in replacement to an existing
application (e.g., specific array location identities are spread around during compression).
However,  many ``bulk  operations'' can   be supported quite naturally, and {\em directly without decompression}.

Here is one use-case: consider the   so-called {\em shallow-water simulation}~\cite{james-sws}.
Suppose one wants to perform these simulations over time at two working precisions, and
obtains two time-series data sets (``two movies''), and also wants to determine the point in time at which the two time-series deviate beyond a threshold.
Such deviations
can be coarsely calculated
using the 
$L_2$-norm distance (which looks at the whole surface)
or 
the Wasserstein distance\footnote{A measure of the least number of edits to make one probability distribution match another.}
that can capture subtle local features (e.g., ``water ripples'').
In this work, we show that a suitably designed lossy compressor can determine both these types of distance metrics with respect to compressed data representations---without first having to decompress.

\paragraph*{Key Contributions}
We now list our two key contributions, providing insights.
First, how
can we incorporate operations 
that act directly in the compressed space? This is achieved by making sure that the compression coefficients are a much smaller proxy to the original data and are obtained through a set of linear operations (detailed in the paper).
Thus one can operate on the coefficients in lieu of the data.
These types of transformations are, in principle, supported by existing compression schemes such as ZFP~\cite{zfp}---but these have not been realized or evaluated in practice.
In this regard, our contribution is to support the ``obvious'' operations such as addition and ``non-obvious'' ones such as Wasserstein distance.
This results in a compressor that does not achieve as high a compression ratio as would have been achievable if we only had aimed for high compression ratios without the need to directly support operations---but with the bonus of having direct operation capability.

Second, even if these compressed-space operations are available, one has to be able to use them meaningfully in actual problem contexts. 
In this regard, we demonstrate these operations in the setting of three distinct applications where the distance between two data sets arises naturally.
We show that it is possible to discern sufficient differences even in the image space of compressed representations.
\footnotemark

\footnotetext{Ideally our study ought to be accompanied by a study of how much we saved in terms of not moving too much data around by operating in the compressed domain.
We plan to embark on this study in our follow-on work.}

\section{Background and Related Work}
\label{sec:background}
% \ggcmt{cite ehrlich and other direct op papers. Also vinu and nithin etc. REF is~\cite{compressed-sensing}}
%

%
%

Lossy data compression methods~\cite{sz,zfp,balazs2017real,kammerl2012real,tang2018real, cusz, cappello-compr, sz3, sz1},
notably 
 ZFP~\cite{zfp} and SZ~\cite{sz}, often achieve compression ratios of 50 or more, while maintaining sufficient fidelity for many applications.
These compressors, developed with scientific data in mind, generally do not make many assumptions regarding the dimensionality of the data or the nature of each dimension. 
Specialized compression algorithms such as MP3~\cite{mp3} for audio, JPEG~\cite{jpeg} for images, or HEVC~\cite{hevc} for video exist, but are outside the scope of our work.
None of these were created with the goal of facilitating operations on compressed data.
\blaz~\cite{blaz} is a lossy compressor for 2-dimensional arrays, allowing certain compressed-space operations such as addition and multiplying by a scalar.
We take inspiration from  \blaz to build a compressor named \pyblaz for higher performance
and also (especially) for direct operation.
%
% Other work such as \cite{amarsinghe_lossless} exists that can perform the compressed-domain transformation on lossless data by using the Lempel Ziv algorithm (LZ77).
%
There are also designs that permit direct operation, but mainly for neural networks~\cite{manipulatingjpeg,uberjpeg,jpegresnet,ehrlich2019deep,compressed-sensing, vinu-arxiv2}.
%

% \ggcmt{Address this comment by Sheng:
% In section 4, the operations supported by PyBlaz are listed such as Negation, addition, doc product, but it's unclear how they are combined with the proposed compressor (proposed in Section 3.1). 

% As a reviewer, I will be more interested in reading how the operation-over-compressed-data is designed and implemented. So, I think this part is better to be put in the main text of the paper. In comparison, I think the definition of the operations i.e, Section 4.1 and 4.2 could be put in the appendix. If the space allows, you can put both in the main text. 
% }
% \tacmt{This has been taken care of}

% \subsection{Scalar Functions on Arrays}
% We implement some common similarity and distance measures in compressed space.
% %
% Specific operations supported
% are
% \textit{$L_2$ norm}~\cite{wiki:norm},  
% %
% \textit{Cosine similarity}~\cite{wiki:cosine_similarity},
% %
% \textit{Structural similarity index measure (SSIM)}~\cite{wiki:structural_similarity},
% %
% % \ggcmt{sz folks show that it is good for scientific data also---emph. that? as this is not for human-oriented compression.}
% %
% %
% \textit{Wasserstein distance}~\cite{wiki:wasserstein_metric},
% all detailed in \S~\ref{subsec:operations} and \S~\ref{subsubsec:approx-ops}.

 \begin{figure*}[tbp]
    \centering
    \includegraphics[width=\textwidth]{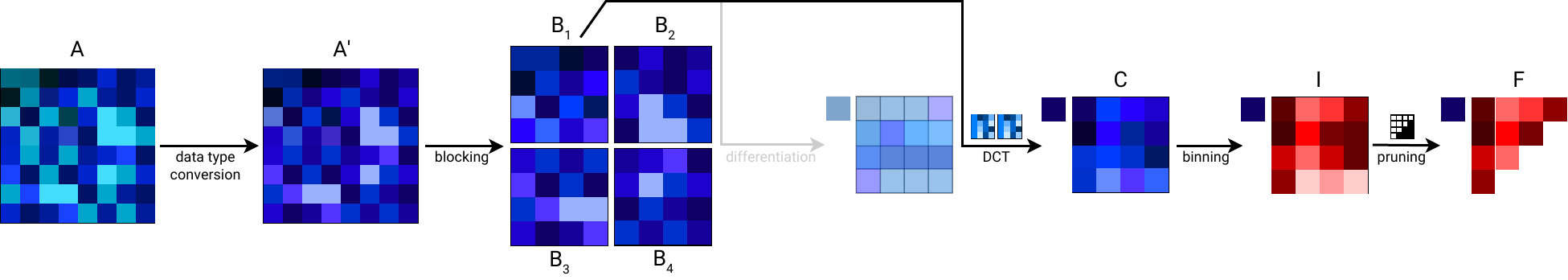} 
    \caption{\pyblaz architecture showing compression of a 2-dimensional array. The blue colors represent floating-point numbers and red colors represent integers. 
    % Each step in the pipeline is marked with a separate name representing the array we are working on in each step. 
    $A$ is the input array, and $A'$ is the array obtained after lowering the precision, $B_{1..4}$ represent the blocks we get after blocking. In this figure, we show the rest of the procedure only on block $B_1$. We perform DCT on each block, after which we obtain an array of coefficients $C$, which is further binned to result in an array of indices $I$. Finally, we apply the pruning mask, shown with black and white colors representing Boolean values, which results in pruned indices represented as $F$. $F$ is later flattened.
    % \hdcmt{Later we use $F$ to refer to this thing flattened. Might be confusing.}
    Unlike in \blaz, we skip the \textit{differentiation} step (called \textit{normalization} in \cite{blaz}), which facilitates certain compressed-space operations explained in detail in \S~\ref{subsec:operations}.}
    %
    % \ggcmt{Mention incoming FP-type,
    % intermediate types etc at a high level when Fig1 is discussed. %
    % Later in expts make sure it is highlighted again.}
   
    %

    % \hdcmt{In the proposed replacement figure, the following appear to be incorrect.\\
    %      $\cdot$ The result of differentiation does not appear to be the derivative of B$_1$.\\
    %      $\cdot$ DCT transform tensor does not appear to decorrelate the second dimension.\\
    %      $\cdot$ C looks the same as B$_1$, and does not look like DCT coefficients.\\
    %      }
    % \tacmt{I think I have upgraded the figure with all the changes that you said. Please feel free to update any changes you feel should be there or let me know. \url{https://drive.google.com/file/d/1zdyAgvBzUQezOoPrP0UxQ0wwbIUtSwjB/view?usp=sharing}}
    \label{fig:compressiondiagram}
\end{figure*}

\subsection{\pyblaz and three related compressors}

We now discuss three closely related compressors to provide some context: ZFP \cite{zfp}, SZ \cite{sz, sz1, sz3, cusz}, and \blaz \cite{blaz}.
We first start with a few concepts common to all compressors, motivating their purpose:
\begin{compactitem}
\item Decorrelation: Decorrelation is any process that reduces the autocorrelation  in data. Orthonormal transformations and prediction are two decorrelation techniques.

\item An orthonormal transformation is a linear transformation that consolidates selectable bands of information while preserving orthogonality and dot products, key to realizing efficient compressed-space operations.

\item  Prediction is describing elements of an array in relation to other nearby elements using a model of the data, such as a sliding template.
\end{compactitem}

\paragraph{ZFP} is a compression method for arbitrary-dimensional \double arrays. 
It begins by blocking input arrays into blocks of size 4 in each direction. 
Each of these blocks is converted to a block floating-point format, such that each block shares the exponent of the biggest element in the block, with other significands scaled appropriately and converted to fix-point form. Next, a near-orthogonal transform is applied to the fixed-point numbers in all directions. The resulting coefficients are converted to negabinary and their bits are encoded in decreasing order of significance using a Huffman-like scheme.

\paragraph{SZ}  is a compression method designed for compressing 1- to 5-dimensional floating-point arrays. 
% SZ compressor can achieve a high compression ratio for compressed data and also maintains a high level of accuracy. 
% The SZ compressor has two stages for compression.
SZ uses a constant, linear, or quadratic prediction model to predict each element in the array based on its neighbors and quantizes the residuals using Huffman coding.  

\paragraph{\blaz}  is a compression method for 2-dimensional \double arrays. \blaz blocks input arrays into $8 \times 8$ blocks. It then saves the first element of each block and encodes the others as the difference from their previous elements. Then, a block-wise discrete cosine transform (DCT) \cite{dct} is applied, resulting in blocks of coefficients. The biggest coefficient in each block is saved, and the others are binned into 255 bins, indexed using 8-bit integers from -127 to 127. Each block of indices is then pruned by 
dropping the $6 \times 6$ square in the higher-index corner of the block
and then flattening the remaining indices.

\subsection{Notations} 
 
 We refer to the length (or size) of an array in each direction as its \textit{shape} and notate shapes using tuples, e.g. (4, 4) for a $4 \times 4$ array, or more generally using lower-case boldface letters (e.g. $\mathbf{x}$), like vectors. Similarly, we notate indices into arrays like vectors, e.g. $\mathbf{i}$ indexes into array $X$ as $X_\mathbf{i}$. Also, $X_{\dots 1}$ is all elements of $X$ whose indices' last coordinate is 1. We use $\sum X$ to notate the sum of all elements in $X$, and $\prod X$ as the product of all elements in $X$. $X + Y$ is element-wise addition of $X$ and $Y$, $X - Y$ is element-wise subtraction of $Y$ from $X$, $X \odot Y$ is the element-wise product of $X$ and $Y$, and $X \oslash Y$ is the element-wise quotient of $X$ and $Y$, broadcasting where necessary. $X^p$ is the element-wise exponentiation of each element of $X$ to the $p$th power and $b^X$ is an array of $b$ exponentiated by elements in $X$. $\lceil X \rceil$ is the ceiling of $X$, an array of integers that are at least as great as their corresponding element of $X$.

% \ggcmt{Say in a recap para that SSIM and Wasserstein track local features better? if so that mini para can be put here.}

% We experiment with different datasets to show the usage of compressed-space operations and different metrics. The purpose of each dataset is to show how and when different operations can be used. We show our work in three different fields. 
% \ggcmt{which 3 fields?}
%
% \ggcmt{This whole section on precision which was here does not belong here. It breaks readability. I've moved it to discussions.}

% \ggcmt{best not to dump extra knowledge in every section that does not flow or fit an argument; dump only what matters. Gunnar connectivity , while cool , is not studied later - so don't dump it... I may nuke it}
% \tacmt{SSIM captures local features whereas Wasserstein is more sensitive for global features}
% \ggcmt{if you want to add that here that might help}
% \hdcmt{I would be careful about general statements about the global and local sensitivity of these metrics. As an example, I could just change the order of a norm to make it more sensitive to local ($L_1$) or global ($L_\infty$) features.}
% \tacmt{I think adding that will again make the readers expectation that we are going to show results related to local and global features. Let's skip it for now.}
% Resolved.

\section{PyBlaz Architecture}
\label{sec:PyBlaz}
% % \tacmt{Modified text so that it looks as original work and not an adaptation}
% As we are interested in a compression method that allows operations on compressed data without decompressing it, we develop our \pyblaz compressor. We take inspiration from \blaz~\cite{blaz}, an existing compressor that incorporates
% similar principles 
% but is less suited to
% handle large arrays. 
% %
% Specifically, \blaz is serial,
% is specialized in dimensionality, and assumes specific input characteristics.
% %
% However, \blaz lent itself to parallelization as well as was
% generalizable to higher dimensions in a relatively straightforward manner.

% \pyblaz is implemented in Python using the PyTorch numerical computation framework with CUDA acceleration.\footnote{\pyblaz already enjoys the performance benefits of GPU-accelerated computing. It is conceivable that one can further optimize PyBlaz more toward its role to support data compression.}
% %
% %
We developed \pyblaz
to compress {\em arbitrary-dimensional} arrays of floating-point numbers.  %
   The key goal
   of this paper
   is, of course, to have \pyblaz \textit{directly} support
   (i.e., without decompression) 
   operations such as the
 dot product, mean, variance, covariance, $L_2$ norm, cosine similarity,  structural similarity, 
 and Wasserstein distance.
 (all detailed
 in \S~\ref{sec:compressedoperations})---all
    without losing compression efficacy.
The compression ratio (\S~\ref{sec:compressionrate}) depends only on compression settings and is independent of data. This is in contrast to other compressors, such as SZ, that might change the compression ratio in order to enforce some $L_\infty$ error bound. Thus, the amount of compression-induced error depends on how closely the settings suit the data being compressed.
 
%

%

%
% \ggcmt{check this - this is hugely important. While this is all cut-up, it is  meant to be one para.}
% \hdcmt{True.}
%

% \ggcmt{check}
% \hdcmt{True.}

% In the following subsections, we describe the \pyblaz compression and decompression process, the compressed-space operations that \pyblaz supports, and the expected compression ratio.

% \ggcmt{Make description of PyBlaz come sooner. Fig1 is already there - start goign thru it.}
% \tacmt{Still making the figure better}
% \ggcmt{Move these notations to Background, Notations}
% \tacmt{Done.}
%

\subsection{Compression Steps}
\label{subsec:Implementation}
% \tacmt{Added a glue statement (first line)}
% PyBlaz consists of two major components, like any other floating-point compressor, compression of the input data and decompression of the compressed data.
%
 
Compression consists of five major steps: data type conversion, blocking, orthonormal transform, binning, and pruning. Each step is explained below and an example of compressing a 2-dimensional array is shown in Figure~\ref{fig:compressiondiagram}. Decompression consists of the compression steps in reverse. As \pyblaz is a lossy compressor, only blocking is exactly invertible. 

%
% \ggcmt{Figure is too gaudy. Make it simpler. If colors serve a purpose, make it clear. Improve layout. Make diagrams neater. Use draw.io which allows interactive editing.
% }

\paragraph{Data type conversion (to lower precision)} 
%
% \ggcmt{Input data rep in many ways. Later expts show the pros/cons. Our framework is parameterized to permit ops
% on any of these types.}
% %
% \ggcmt{Coeff fixpt width alloc. See if relevant to bring it in here.}
% %
% \tacmt{Changed text, please read}
Uncompressed scientific data is often found in 64-bit floating-point (\double) arrays. During compression, it is sometimes advantageous to convert the array elements to a lower-precision data type because the upcoming binning and pruning steps will further discard information. Lowering the precision not only reduces the compressed form memory usage, but also speeds up compression, decompression, and certain compressed-space operations. \pyblaz supports Brain Floating Point (\bfloat) \cite{bfloat16}, \half, \single, and \double as the precision types to which one can convert the input data. Loss is expected when converting to a type with fewer significand bits or converting to a type with insufficient dynamic range.

\paragraph{Blocking}
Blocking allows subsequent steps in the compression process to be performed on each block independently, facilitating parallelization.
%
% Blocking also keeps the size of the blocks in each direction a power of two to enable efficient decorrelation from the orthonormal transform. 
%
\pyblaz supports all block shapes that are a power of two in all directions, including shapes that are not the same length in all directions, i.e. non-hypercubic shapes. During blocking, an input array shaped $\mathbf{s}$, with dimensionality $d = |\mathbf{s}|$, is padded with zeros such that its size in each direction is a multiple of the block size in the corresponding direction. Let the block shape be $\mathbf{i}$ (the maximum \textit{intrablock} index), and let the arrangement of blocks in the reshaped array have shape $\mathbf{b} = \lceil \mathbf{s \oslash i} \rceil$ (the maximum \textit{block} index). Then the reshaped array will have shape $\mathbf{bi}$.
% \tacmt{Added back the footnote to the main text. Footnote destroys the flow of paper. Also, we do explain something similar later in compression rate regarding the size of array, so why not have everything here and not as footnote.}
For example, using an input array shape \mbox{(3, 224, 224)} and block shape \mbox{(4, 4, 4)}, the reshaped array will have shape \mbox{(1, 56, 56, 4, 4, 4)}.

\paragraph{Orthonormal transform}
Each block is then transformed into coefficients of some orthonormal transform, which we use to reduce periodic (wave-like) redundancy by consolidating data of the same spatial frequency into a scalar coefficient. \pyblaz uses DCT \cite{dct} by default, such that the DCT coefficients correspond to the correlation between the block and sampled cosine functions of different frequencies in different directions. Let $\{\mathbf{H}_1, \dots, \mathbf{H}_{d} \}$ be the DCT matrices for each block size. Matrix $\mathbf{H}$ for block size $s$ contains each $i$th element of each $j$-frequency sampled cosine function such that 
$\mathbf{H}_{ij} = \sqrt{\frac{1 + (j>1)}{s}} \cos \frac{ \pi i (2j+1)}{2s}$, where $i \in [1..s]$ and $j \in [1..s]$.
Let $\Sigma$ be a list of at least $2d$ symbols to use as indices. Then if $B$ is the blocked array, the block-wise transformed array is 
$C_{[\mathbf{1}..\mathbf{b}] [\Sigma_{d+1}..\Sigma_{2d}]} =
B_{[\mathbf{1}..\mathbf{b}] [\Sigma_{1}..\Sigma_{d}]} 
\mathbf{H}_{1 \; \Sigma_{1} \Sigma_{1+d}} 
\mathbf{H}_{2 \; \Sigma_{2} \Sigma_{2+d}}
\dots
\mathbf{H}_{d \; \Sigma_{d} \Sigma_{2d}}$
using Einstein's summation notation~\cite{einsum}. 
An example is explained in Appendix~\ref{sec:dctexample}.
\pyblaz also supports other transforms besides DCT, such as the Haar wavelet transform. The block can be recovered, modulo floating-point rounding error, by multiplying these coefficients with their corresponding sampled functions and summing them. 

\paragraph{Binning} 
Binning coarsens the coefficient space, allowing coefficients to be referred to using shorter descriptors (fewer bits). To perform binning, the biggest elements per block of the coefficients array $C$ are collected as $N_\mathbf{k} = \lVert C_{\mathbf{k}} \rVert_\infty$ for all block indices $\mathbf{k}$. Then the elements in $C$ are binned into a number of bins according to the bin index type, an integer type. \pyblaz supports \byte, \short, \inttt, and \intsf. The number of bins is the number of values distinguishable by the data type minus one, and more bins enables finer rounding. The values of the bins in each block are centered at 0 and range to the biggest element. We define an index type radius $r = 2^{b-1}-1$, where $b$ is the number of bits in the integer type.
The values are binned by multiplying them by $r$, dividing by the biggest coefficient in their blocks, and rounding to the nearest integer.
The result is an array of indices $I_{\mathbf{k}} = \text{int}( \text{round}( rC_{\mathbf{k}} \oslash N_\mathbf{k}))$ for all $\mathbf{k}$.

\paragraph{Pruning}
 Pruning allows selection of certain coefficient indices, and thus certain frequencies, to keep in the compressed array representation. The choice of indices to keep depends on the application. For example, noise of a certain frequency may be filtered by pruning its corresponding coefficient. Indices to keep are specified using a pruning mask $P$, shaped $\mathbf{i}$. Loss in this step stems from the dropping indices, as it is effectively rounding to zero. After pruning, the remaining indices are flattened into a sequence $F_\mathbf{k}$ for all blocks $\mathbf{k}$. Because $P$ is saved in the compressed representation, $F$ can be unflattened with zeros in place of unspecified indices.
%
% \tacmt{Move decompression here, instead on being in the first paragraph}

% \ggcmt{At the end of this pipeline we are left with 'normalized coeff'. 
% Flattened seq.
% 1D.
% We also have the biggest coeff for un-normalizing.
% Orig shape; mask to prune coeff. 
% Block-size..
% }
% \tacmt{Explained what we are left after compression, and information of the set that we get}

\subsection{Compressed Form}

The result of compressing the input data is a compressed array consisting of the original shape $\mathbf{s}$, the block shape $\mathbf{i}$, the biggest coefficient for each block $N$, the flattened specified bin indices $F$, as well as information that is required for decompression, such as the pruning mask. All components and their memory usage are listed in \S~\ref{sec:compressionrate}. In our discussion of compressed-space operations (\S~\ref{sec:compressedoperations}), we consider only the set $\{\mathbf{s}, \mathbf{i}, N, F\}$ as a simplification of a compressed array. From this, an array can be decompressed by scaling $F$ by $N$ appropriately, unflattening blocks, performing the inverse orthonormal transform, merging blocks, and cropping to the original shape.

% \tacmt{Can be removed and added to FMCAD. A short discussion is also given in the introduction section}

% \textit{\textcolor{red}{The following features are added to \pyblaz in comparison to \blaz:
% \begin{itemize}
%     \item Arbitrary dimensionality: \blaz supports only 2-dimensional arrays.
%     \item Variable bin index type: int8, int16, int32, int64 can be used in \pyblaz as compared to \blaz which uses only int8.
%     \item Multiple data types: \pyblaz supports using bfloat16, float16, float32 and float64 for internal computations and for storing certain compressed array elements; \blaz uses only float64.
%     \item Variable block shapes: \blaz is limited to the $8\times 8$ square blocks. \pyblaz can have a greater variety of block shapes.
%     \item Variable invertible transform functions: Blaz uses only DCT. \pyblaz also uses DCT by default, but different transforms can be substituted. Note that certain operations such as variance assume orthonormality. 
%     \item CUDA acceleration: The tool can use GPUs for parallel processing of multiple blocks created in the blocking step. This accelerates compression, decompression, and compressed-space operations on large arrays.
% \end{itemize}}}

\section{Compressed-Space Operations}
\label{sec:compressedoperations}
%

% We now detail how
%  \pyblaz
% directly supports
% compressed-space operations
% on data that has been compressed to a significant degree 
% while also aiming for
% high
% computational efficiency.
% %
% Given the compression steps described in \S~\ref{subsec:Implementation}, there are certain operations that are efficient to perform.

% The compressed-space operations are mathematically equivalent to when decompression  is performed on compressed data and then the operations are performed  on decompressed arrays, but because \pyblaz is a lossy compressor, the operations may carry on loss that was introduced during compression or introduce additional loss, which exhibits as error compared to operations on arrays that were never compressed. Such error depends on compression settings, and our experiments (\S~\ref{sec:experimentsandresults}) explore their effect using various datasets.

% \hdcmt{Added the following note about differentiation.}
Table~\ref{table:Operations} lists
all the compressed-space operations along with the result type  and source of error that can occur in each operation. All of the operations, except the approximate Wasserstein distance, are differentiable. This feature enables their incorporation into gradient-based optimization pipelines, such as those that form the backbone of state-of-the-art ML models. We will explain these operations with pseudocode.
%
% \tacmt{Added a glue statement here.}

% \ggcmt{
% In all the compr space ops,
% arrays are the input
% (captured by the signature
% (s,i,N,F)...)
% and the output is an
% array or a scalar quantity
% such as ... and ... . 
% }

\begin{table}
    \centering
    \renewcommand{\arraystretch}{1.5}
\begin{tabular}{|l|l|l|}
\hline
    \textbf{Operation} & 
    \textbf{\shortstack[l]{Result Type}} & \textbf{\shortstack[l]{\shortstack[l]{Source\\ of \\Error}}}\\
    \hline
    \hline
    Negation & $\mathrm{(Array)\; } \{\mathbf{s}, \mathbf{i}, N, -F\}$ & none \\
    \hline
    \shortstack[l]{Element-wise\\addition}& $\mathrm{(Array)\; }\{\mathbf{s}, \mathbf{i}, N, F\}\; $& 
    \shortstack[l]{rebinning}\\
    \hline
    \shortstack[l]{Addition of \\ a scalar} & $\mathrm{(Array)\; }\{\mathbf{s}, \mathbf{i}, N, F\}$ & rebinning\\
    \hline
    \shortstack[l]{Multiplication \\ by a scalar} & $\mathrm{(Array)\; }\{\mathbf{s}, \mathbf{i}, N \odot |x|, F \odot \mathrm{sign}(x)\}$ & none \\
    \hline
    Dot product & $\mathrm{(Scalar)\; }\sum (\hat{C}_1 \odot \hat{C}_2)$ & none\\
    \hline
    Mean & $ \mathrm{(Scalar)\; }\mathrm{mean}(\hat{C}_{\dots 1}) \oslash \prod (\mathbf{i}^\frac{1}{2})$& none\\
    \hline
    Covariance & $\mathrm{(Scalar)\; }\mathrm{mean}(\hat{C}_1 \odot \hat{C}_2)$ & none\\
    \hline
    Variance & $\mathrm{(Scalar)\; }$ Covariance($A,A$) & none\\
    \hline
    $L_2$ norm & $\mathrm{(Scalar)\; }$$\lVert \hat{C} \rVert_2$ & none\\
    \hline
    Cosine similarity & $\mathrm{(Scalar)\; }$$\dfrac{p}{m}$&none \\
    \hline
    SSIM & $\mathrm{(Scalar)\; }$$ l^{w_l} c^{w_c} s^{w_s}$& none\\
    \hline
\shortstack[l]{Approximate\\
Wasserstein\\ distance} &(Scalar)\;  $\left ( \dfrac{ \sum |P_{A'}-P_{B'}|^p }{\prod \lceil \mathbf{s} \oslash \mathbf{i} \rceil} \right )^\frac{1}{p}$& 
\shortstack[l]{Error as a \\ function of \\ block size}\\
    \hline
    
\end{tabular}
\caption{The list of operations supported by \pyblaz along with their result types  (array of scalar).
 Some details are explained in
 \S~\ref{subsec:operations} and \S~\ref{subsubsec:approx-ops}. 
 Notably, most of the compressed-space operations 
introduce no additional error (beyond the inevitable error incurred during compression.)
}
\label{table:Operations}
\end{table}

% \ggcmt{Define an ADT or just a tuple of elements that is the space over which the ops are done and the results are delivered.
% Then in the algos,
% refer to the algos.}
% \tacmt{This is explained just before we started talking about compressed space operations, it will be redundant to add that information here, which we define just one paragraph before and slightly just after} 

\subsection{Operations in \pyblaz}
\label{subsec:operations}
The output obtained from compressing the input data is used to perform the compressed-space operations. As discussed before, we use the set $\{\mathbf{s}, \mathbf{i}, N, F\}$ as a simplification of a compressed array.
The algorithms presented in this section
employ several element-wise array
operations which have a natural
implementation on GPUs (facilitated
by our use of GPU-based PyTorch in realizing
\pyblaz).
The compression pipeline facilitates these operations due to two key properties. (1) Each block of $F$ is proportional to each block of transform coefficients, and scaling $F$ by $N$ recovers the specified coefficients. Thus, operating on $F$ and in some cases $N$ is equivalent to operating on the coefficients. 
(2) Because the orthonormal transform preserves orthonormality and dot products, the inverse transform is unnecessary for addition and multiplication and for obtaining related summative information, such as magnitudes and variances.
%

% \ggcmt{Ops supported

% Blaz was aimed at 2D and support add/sup/matrix ops

% We had to go in a diff direction by skipping diff etc..}

% % Many compressed-space operations were already available in \blaz.  Because we skipped the \textit{differentiation} step in \blaz, our implementation of these operations differs. 

% \ggcmt{TABLE of Ops, approx or not, delivered result-type : a table will be good.}
% \hdcmt{Since there is only one approximate operation, approximation need not have a column.
% \tacmt{Makes sense}}

% \ggcmt{e.g. say These ops represent those that one likes to have in standard libraries, plus ops that support distance measures between two compr arrays. Some of these are in the expts and some planned for future work.}
% \tacmt{This table has moved above.}

\subsubsection{Negation}
\label{sec:negation}
Negation (Algorithm~\ref{alg:negation}) is done by negating $F$. Because the bin indices are proportional to the coefficients, negating the bin indices is equivalent to negating the coefficients that would result during decompression.

\begin{algorithm}
    \caption{Negation($A$)}\label{alg:negation}
    \KwData{compressed array $A =\{\mathbf{s}, \mathbf{i}, N, F\}$}
    \KwResult{the negated array}
    \Return{$\{\mathbf{s}, \mathbf{i}, N, -F\}$}\;
\end{algorithm}
\subsubsection{Element-wise addition}
\label{sec:addition}
To add two compressed arrays (Algorithm~\ref{alg:addition}), we calculate a sum of their specified coefficients and scale the indices to the possibly changed biggest coefficients.
% \footnote{The $L_\infty$ norm step is differentiable because, given an array $X$, the derivative of its infinity norm $\lVert X \rVert_\infty$ is an array in the same shape as $X$ with 1 in the element corresponding to the biggest element in $X$ and 0 elsewhere.}

\begin{algorithm}
    \caption{Addition($A,B$)}\label{alg:addition}
    \KwData{compressed arrays $A = \{\mathbf{s}, \mathbf{i}, N_1, F_1\}, B = \{\mathbf{s}, \mathbf{i}, N_2, F_2\}$}
    \KwResult{the element-wise sum of $A$ and $B$}
    $b \gets$ the number of bits in the element type of $F$\;
    $r \gets 2^{b-1}-1$\;
    $\hat{C} \gets F_1 \odot N_1 + F_2 \odot N_2$\;
    $N_\mathbf{k} \gets \lVert \hat{C}_\mathbf{k} \rVert_\infty \oslash r: \; \mathbf{k} \in [\mathbf{1}..\lceil \mathbf{s} \oslash \mathbf{i} \rceil]$\;
    $F \gets \mathrm{int}(\mathrm{round}(\hat{C} \oslash N))$\;
    \Return{$\{\mathbf{s}, \mathbf{i}, N, F\}$}\;
\end{algorithm}

\subsubsection{Addition of a scalar}
This and some later operations depend on the specified coefficients (Algorithm~\ref{alg:coefficients}).

\begin{algorithm}
    \caption{SpecifiedCoefficients($A$)}\label{alg:coefficients}
    \KwData{compressed array $A = \{\mathbf{s}, \mathbf{i}, N, F\}$}
    \KwResult{specified coefficients of a compressed array}
    $b \gets$ the number of bits in the element type of $F$\;
    $r \gets 2^{b-1}-1$\;
    \Return{$N \odot F \oslash r$}\;
\end{algorithm}

If the block shape is $\mathbf{i}$ and the first coefficients in each block were not pruned away, the first coefficient in each block is the mean of the uncompressed block scaled by $c = \prod (\mathbf{i}^\frac{1}{2})$. Then the mean of the uncompressed array is the mean of all first coefficients divided by $c$. The addition of a scalar (Algorithm~\ref{alg:scalaraddition}) is then achieved by adding the scalar multiplied by $c$ to all first coefficients.
% \tacmt{Instead of writing Agorithm 3, 4 5... where they are used in other algorithms it is better to write the name and then mark it using algorithm commnts as to which algorithm we want to refer to. Giving the name adds the ease to understand the algorithm easily by just understanding the what the name of the algorithm implies. I have corrected this in algorithms, please check}
% \hdcmt{I shortened the comments.}
\begin{algorithm}
    \caption{AdditionOfScalar($A,x$)}\label{alg:scalaraddition}
    \KwData{compressed array $A =\{\mathbf{s}, \mathbf{i}, N, F\}$, scalar $x$}
    \KwResult{the array with $x$ added to all elements}

    $\hat{C} \gets$  SpecifiedCoefficients(A); \hfill // Algorithm~\ref{alg:coefficients}\\
    $N_\mathbf{k} \gets \lVert \hat{C}_\mathbf{k} \rVert_\infty: \; \mathbf{k} \in [\mathbf{1}..\lceil \mathbf{s} \oslash \mathbf{i} \rceil]$ \;
    $\hat{C}_{\dots 1} \gets \hat{C}_{\dots 1} + x\prod (\mathbf{i}^\frac{1}{2})$\;
    $b \gets$ the number of bits in the type of $F$\;
    $r \gets 2^{b-1}-1$\;
    $F \gets \mathrm{int}( \mathrm{round}( \hat{C} \odot r \oslash N))$
   
    \Return{$\{\mathbf{s}, \mathbf{i}, N, F\}$}\;
\end{algorithm}

\subsubsection{Multiplication by a scalar}
\label{sec:scalarmultiplication}
Multiplication by a scalar (Algorithm~\ref{alg:scalarmultiplication}) is done by multiplying $N$ by the absolute value of the scalar, and if the scalar is negative, negating $F$.

\begin{algorithm}
    \caption{MultiplicationByScalar($A,x$)}\label{alg:scalarmultiplication}
    \KwData{compressed array $A=\{\mathbf{s}, \mathbf{i}, N, F\}$, scalar $x$}
    \KwResult{the array multiplied by $x$}
    \Return{$\{\mathbf{s}, \mathbf{i}, N \odot |x|, F \odot \mathrm{sign}(x)\}$}\;
\end{algorithm}

In addition to the above operations, some other operations are facilitated by skipping the differentiation step.
\subsubsection{Dot product}
\label{sec:dot}
The dot products before and after an orthonormal transform are equal. Thus, the compressed-space dot product (Algorithm~\ref{alg:dot}) is the dot product of specified coefficients.

\begin{algorithm}
    \caption{DotProduct($A,B$)}\label{alg:dot}
    \KwData{compressed arrays $A = \{\mathbf{s}, \mathbf{i}, N_1, F_1\}, B = \{\mathbf{s}, \mathbf{i}, N_2, F_2\}$}
    \KwResult{the dot product of $A$ and $B$}
    $\hat{C}_1 \gets$ SpecifiedCoefficients(A); \hfill // Algorithm~\ref{alg:coefficients} \\
    $\hat{C}_2 \gets$  SpecifiedCoefficients(B)\;
    \Return{$\sum (\hat{C}_1 \odot \hat{C}_2)$}\;
\end{algorithm}

\subsubsection{Mean}
\label{sec:mean}
Because the first coefficient in each block is the uncompressed block mean scaled by $c$, the mean (Algorithm~\ref{alg:mean}) is obtained by averaging the first coefficients in all blocks and dividing this average by $c$. The block-wise mean is obtained by collecting all the first coefficients divided by $c$, i.e. $\hat{C}_{\dots 1} \oslash c$.

\begin{algorithm}
    \caption{Mean($A$)}\label{alg:mean}
    \KwData{compressed array $A = \{\mathbf{s}, \mathbf{i}, N, F\}$}
    \KwResult{the mean of the array}
    $\hat{C} \gets$ SpecifiedCoefficients(A); \hfill // Algorithm~\ref{alg:coefficients}\\
    \Return{$ \mathrm{mean}(\hat{C}_{\dots 1}) \oslash \prod (\mathbf{i}^\frac{1}{2})$}\;
\end{algorithm}

\subsubsection{Covariance}
\label{sec:covariance}
Covariance (Algorithm~\ref{alg:covariance}) is obtained by getting the mean of the element-wise product of the centered coefficients of two compressed arrays \cite{jpegresnet}. Centered coefficients represent an uncompressed array with its mean subtracted. Block-wise covariance is also available by getting the block-wise means of this product.

\begin{algorithm}
    \caption{Covariance($A,B$)}\label{alg:covariance}
    \KwData{compressed arrays $A = \{\mathbf{s}, \mathbf{i}, N_1, F_1\}, B = \{\mathbf{s}, \mathbf{i}, N_2, F_2\}$}
    \KwResult{the covariance of $A$ and $B$}
    $\hat{C}_1 \gets$ SpecifiedCoefficients(A); \hfill // Algorithm~\ref{alg:coefficients}\\
    $\hat{C}_2 \gets$  SpecifiedCoefficients(B)\;
    $c \gets \prod \lceil \mathbf{s} \oslash \mathbf{i} \rceil$\;
    $\hat{C}_{1 \;\; \dots 1} \gets \hat{C}_{1 \;\; \dots 1} - (\sum \hat{C}_{1 \;\; \dots 1}) \oslash c$\;
    $\hat{C}_{2 \;\; \dots 1} \gets \hat{C}_{2 \;\; \dots 1} - (\sum \hat{C}_{2 \;\; \dots 1}) \oslash c$\;
    \Return{$\mathrm{mean}(\hat{C}_1 \odot \hat{C}_2)$}\;
\end{algorithm}

\subsubsection{Variance}
\label{sec:variance}
Variance (Algorithm~\ref{alg:variance}) is obtained by finding the covariance of an array with itself. Block-wise variance is available as the block-wise mean of squared block-wise centered coefficients. Standard deviation or block-wise standard deviation are available as the square root of variance or block-wise variance.

\begin{algorithm}
    \caption{Variance($A$)}\label{alg:variance}
    \KwData{compressed array $A =\{\mathbf{s}, \mathbf{i}, N, F\}$}
    \KwResult{the variance of the array}
    $v \gets$ Covariance($A,A$); \hfill // Algorithm~\ref{alg:covariance}\\
    \Return{$v$}\;
\end{algorithm}

\subsubsection{$L_2$ norm}
The $L_2$ norm \cite{wiki:norm}, also known as Euclidean norm, is commonly used to measure physical or spatially meaningful distances. Taking advantage of orthonormality, we can obtain the $L_2$ norm by calculating the $L_2$ norm of the specified coefficients (Algorithm~\ref{alg:norm2}).

\begin{algorithm}
    \caption{$L_2$Norm($A$)}\label{alg:norm2}
    \KwData{compressed array $A =\{\mathbf{s}, \mathbf{i}, N, F\}$}
    \KwResult{the $L_2$ norm of the array}
    $\hat{C} \gets$ SpecifiedCoefficients($A$); \hfill // Algorithm~\ref{alg:coefficients}\\
    \Return{$\lVert \hat{C} \rVert_2$}\;
\end{algorithm}

\paragraph{Cosine similarity} \cite{wiki:cosine_similarity} captures the degree of angular similarity between two vectors. We obtain cosine similarity by dividing the dot product of two arrays by the product of their $L_2$ norms (Algorithm~\ref{alg:cosinesimilarity}).

\begin{algorithm}
    \caption{CosineSimilarity($A,B$)}\label{alg:cosinesimilarity}
    \KwData{compressed arrays $A = \{\mathbf{s}, \mathbf{i}, N_1, F_1\}, B = \{\mathbf{s}, \mathbf{i}, N_2, F_2\}$}
    \KwResult{the cosine similarity between $A$ and $B$}
    $p \gets$ DotProduct($A,B$); \hfill // Algorithm~\ref{alg:dot}\\
    $m \gets$ $L_2$Norm($A$) * $L_2$Norm($B$) \hfill // Algorithm~\ref{alg:norm2}\\
    \Return{$\dfrac{p}{m}$}\;
\end{algorithm}

\subsubsection{Structural similarity index measure (SSIM)}
\label{sec:ssim}
SSIM \cite{wiki:structural_similarity} was developed to assess the visual similarity of images to humans, but has also been used to evaluate the quality of compressed data~\cite{cappello-compr}. SSIM compares the luminance, contrast, and structure of the images to assign a similarity score from 0 to 1 through a weighted product of these three terms (Algorithm~\ref{alg:ssim}).

\begin{algorithm}
    \caption{StructuralSimilarityIndexMeasure($A,B$)}\label{alg:ssim}
    \KwData{compressed arrays $A = \{\mathbf{s}, \mathbf{i}, N_1, F_1\}, B = \{\mathbf{s}, \mathbf{i}, N_2, F_2\}$, luminance stabilizer $s_l$, contrast stabilizer $s_c$, luminance weight $w_l$, contrast weight $w_c$, structure weight $w_s$}
    \KwResult{the SSIM between $A$ and $B$}
    $\mu_{A} \gets$ Mean($A$); \hfill // Algorithm~\ref{alg:mean}\\
    $\mu_{B} \gets$ Mean($B$)\;
    $\sigma^2_{A} \gets$ Variance($A$); \hfill // Algorithm~\ref{alg:variance}\\
    $\sigma^2_{B} \gets$ Variance($B$)\;
    $\sigma_{A} \gets \sqrt{\sigma^2_{A}}$\;
    $\sigma_{B} \gets \sqrt{\sigma^2_{B}}$\;
    $\sigma_{AB} \gets$ Covariance($A,B$); \hfill // Algorithm~\ref{alg:covariance}\\

    $l \gets \dfrac{2\mu_{A}\mu_{B} + s_l}{\mu_{A}^2+\mu_{B}^2+s_l}$\;
    $c \gets \dfrac{2\sigma_{A}\sigma_{B}+s_c}{\sigma_{A}^2+\sigma_{B}^2+s_c}$\;
    $s \gets \dfrac{\sigma_{AB}+\frac{s_c}{2}}{\sigma_{A}\sigma_{B}+\frac{s_c}{2}}$\;

    \Return{$l^{w_l} c^{w_c} s^{w_s}$}\;
\end{algorithm}

% \ggcmt{Try to end each such op description
% with algo pseudo-code that shows 
% how it is realized using the 
% output of the Fig-1 pipeline.
% Take output vector of fig-1 
% and act on it to produce a 
% quantity which is this op's output}

\subsection{Approximate operations in \pyblaz}
\label{subsubsec:approx-ops}

% \ggcmt{check this glue}
% \hdcmt{True.}
While we have the possibility of realizing many approximate operations in \pyblaz, we discuss only one that is relevant to this work, namely the Wasserstein distance~\cite{wiki:wasserstein_metric}, used to measure the distance between probability distributions. Because it can be described as the cost of transforming one distribution to another by moving probability mass or density, it is also known as the Earth mover's distance. 
%Wasserstein distance is used to generate smooth interpolation between probability distributions and can manipulate probability distributions with disjoint support. 
% Such distributions are sometimes challenging for other distance metrics such as Euclidean distance.
It is especially useful where one distribution is a perturbed version of another.

In \pyblaz, we can use the block-wise mean to find approximations of arbitrary operations on uncompressed arrays. The granularity of the approximation depends on the block shape specified for the blocking step during compression. At one extreme, one-element blocks will result in an exact operation while discarding all compression benefits. Thus, the choice of block shape should anticipate necessary approximate operations.
\subsubsection{Approximate Wasserstein distance}
\label{sec:approximatewasserstein}
% Let the uncompressed arrays $A$ and $B$, both of shape $\mathbf{s}$,
% be compressed with block shape $\mathbf{i}$. 
% % We obtain block-wise means of the data which is passed through a softmax function. This function converts the obtained block-wise means of the array to a probability distribution.  
% Then their block-wise means will be $A'$ and $B'$, both of shape $\lceil \mathbf{s} \oslash \mathbf{i} \rceil$.
% We then sort them and name these sorted arrays $P_{A'}$ and $P_{B'}$ respectively. Then we approximate the $p$-order Wasserstein distance between $A$ and $B$ as $\left ( \frac{1}{\prod \lceil \mathbf{s} \oslash \mathbf{b} \rceil} \sum |P_{A'}-P_{B'}|^p \right )^\frac{1}{p}$.
By using the block-wise mean, we can obtain an approximation of decompressed arrays, and thus can perform the conventional $p$-order Wasserstein distance between them (Algorithm~\ref{alg:approximatewasserstein}). Because the Wasserstein distance measures distances between probability distributions, we ensure that the arrays are probability distributions by applying the softmax function\footnote{$\mathrm{softmax}(X) = \frac{e^{X}}{\sum e^X}$, where $e$ is the base of the natural logarithm.} to them if they are not.
% We actually softmax regardless, and should not.
Because sorting is involved, this function is not differentiable.

\begin{algorithm}
    \caption{ApproximateWassersteinDistance($A,B$)}\label{alg:approximatewasserstein}
    \KwData{compressed arrays $A = \{\mathbf{s}, \mathbf{i}, N_1, F_1\}, B = \{\mathbf{s}, \mathbf{i}, N_2, F_2\}$, order $p$}
    \KwResult{the approximate Wasserstein distance between $A$ and $B$}
    $c \gets \prod{(\mathbf{i}^\frac{1}{2})}$\;
    $A' \gets$ SpecifiedCoefficients($A$)$_{\dots 1} \oslash c$; \hfill // Algorithm~\ref{alg:coefficients}\\
    $B' \gets$ SpecifiedCoefficients($B$)$_{\dots 1} \oslash c$\;
    \If{$\sum A' \neq 1$}
    {
        $A' \gets \mathrm{softmax}(A')$\;
    }
    \If{$\sum B' \neq 1$}
    {
        $B' \gets \mathrm{softmax}(B')$\;
    }
    $P_{A'} \gets \mathrm{sorted}(A')$\;
    $P_{B'} \gets \mathrm{sorted}(B')$\;
    
    \Return{$\left ( \dfrac{ \sum |P_{A'}-P_{B'}|^p }{\prod \lceil \mathbf{s} \oslash \mathbf{i} \rceil} \right )^\frac{1}{p}$}\;
\end{algorithm}

\vspace{1ex}
\noindent We now define a crucial figure of merit of a compressor, namely the compression ratio it delivers.

\subsection{Compression Ratio}
\label{sec:compressionrate}

% \ggcmt{Compression ratio and why it matters and how we derive it, and what influences it the most}
% \tacmt{Done. Added why compression ratio needs to be calculated and checked the text for how we dervie it and what influences it the most.}
% \tacmt{Moving this section after compressed space operation, as it better goes with the flow of paper. We have in the previous section obtained the set of stuff we get from compression and then we uses that information heavily in compressed space operations so it makes more sense to have that information here and then we can talk about compression ratio}
% Lossy compressors are used because of their high compression ratio and gives an indication of how effective the compression technique is, hence to identify the effectiveness of  \pyblaz it is important to find its compression ratio. 

% The compression ratio is determined independently of data. 
After compressing an array whose original shape is $\mathbf{s}$, with dimensionality $d=|\mathbf{s}|$, using block shape $\mathbf{i}$, $f$-bit floating point numbers, $i$-bit integers as the bin index type, and a pruning mask $P$,
certain quantities are to be stored. Because $P$ and $N$ have shapes $\mathbf{i}$ and $\lceil \mathbf{s} \oslash \mathbf{i} \rceil$ respectively, their shapes need not be stored, only their elements flattened into a 1-dimensional sequence. 
The components to be stored are
\begin{compactitem}
    \item the floating point and integer types, specified in $4$ bits,
    \item $\mathbf{s}$, which takes $64d$ bits,
    \item a marker for the end of $\mathbf{s}$, taking up to $64$ bits,
    \item $\mathbf{i}$, which takes $64d$ bits,
    \item $P$ flattened, which takes $\prod \mathbf{i}$ bits,
    \item $N$ flattened, taking $f \prod \lceil \mathbf{s} \oslash \mathbf{i} \rceil$ bits, and
    \item $F$ taking $i (\sum P) (\prod \lceil \mathbf{s} \oslash \mathbf{i} \rceil )$ bits.
\end{compactitem}

Thus, if the input array were provided with $u$-bit elements, the compression ratio approaches 
% \hdcmt{All big deal equations like this should be numbered. I recommend not making a big deal about it because this is the only big deal equation in the paper.}
\begin{equation*}
    \frac{u \prod \mathbf{s}}
{(f + i \sum P) \prod \lceil \mathbf{s} \oslash \mathbf{i} \rceil}
\end{equation*}

as array size approaches infinity. 

The compression settings that most impact the compression ratio are the bin index type and pruning mask.
For example, with an input array shaped (3, 224, 224) of 64-bit elements, block shape (4, 4, 4), floating-point type \single, index type \short, and no pruning, the compression ratio will be $\approx$ 2.91. Using \byte and pruning half the indices, the compression ratio will be $\approx$10.66.

\subsection{Compression Error}
\label{subsec:Error}
 
Compression error is introduced in the data type conversion, orthonormal transform, binning, and pruning steps. Error in the data type conversion and orthonormal transform steps consists only of floating-point rounding error, which we will exclude from this discussion. Here we describe error in the other steps.

In the \textbf{binning} step, we will consider the maximum error per block, which is half the width of a bin, determined by the highest-magnitude coefficient in the block and the number of bins. If the magnitude of the biggest coefficient in block $\mathbf{k}$ is $N_\mathbf{k}$, then the range of numbers that the bins must cover is $2 N_\mathbf{k}$. Recall that because the bins are centered at zero, the number of bins given an integer bin index type is the number of values distinguishable by that type minus one, which is also twice the radius $r$ plus one, so we have $2r + 1$ bins. Then, the width of a bin is $\frac{2 N_\mathbf{k}}{2r + 1}$, and so the maximum error in terms of coefficients is $\frac{ N_\mathbf{k}}{2r + 1}$.

In the \textbf{pruning} step, error is introduced by pruning non-zero bin indices, which correspond to non-zero coefficients. Therefore, in terms of coefficients, the error is simply the corresponding coefficients of the pruned indices.

As for how coefficient errors translate to errors in the uncompressed space, we must consider the orthonormal transform used. Recall that the basis vectors in orthonormal transform matrices are of unit length. Therefore, if a coefficient were to maximally influence an element of the decompressed array, the basis vector should consist of all zeros except for a one in some element, such as if we used the standard basis as a transform matrix (although doing so is useless). Therefore, by changing a single coefficient, the maximum error that may appear in an element of a decompressed array (i.e. the L$_\infty$ error bound) is the same as the error in the coefficient.

Of course, binning and pruning usually change more than one coefficient. Their combined effects on one element of a decompressed array is the sum of the corresponding elements of the transform matrix scaled by the coefficients. Therefore, if $C$ is the coefficients and $\mathbf{i}$ is the block shape, then in block $\mathbf{k}$, the only L$_\infty$ error bound we can provide is a rather loose $\Vert C_\mathbf{k} \Vert_\infty \prod{\mathbf{i}}$.
However, due to the dot product-preserving properties of orthonormal transforms, we know the the L$_2$ error in the block is the same as the L$_2$ norm of the coefficient errors in the block.
\subsection{Compression, Decompression, and Operation Time}
\label{subsec:Time}

\begin{figure}
    \centering
    \includegraphics[width=0.48\textwidth]{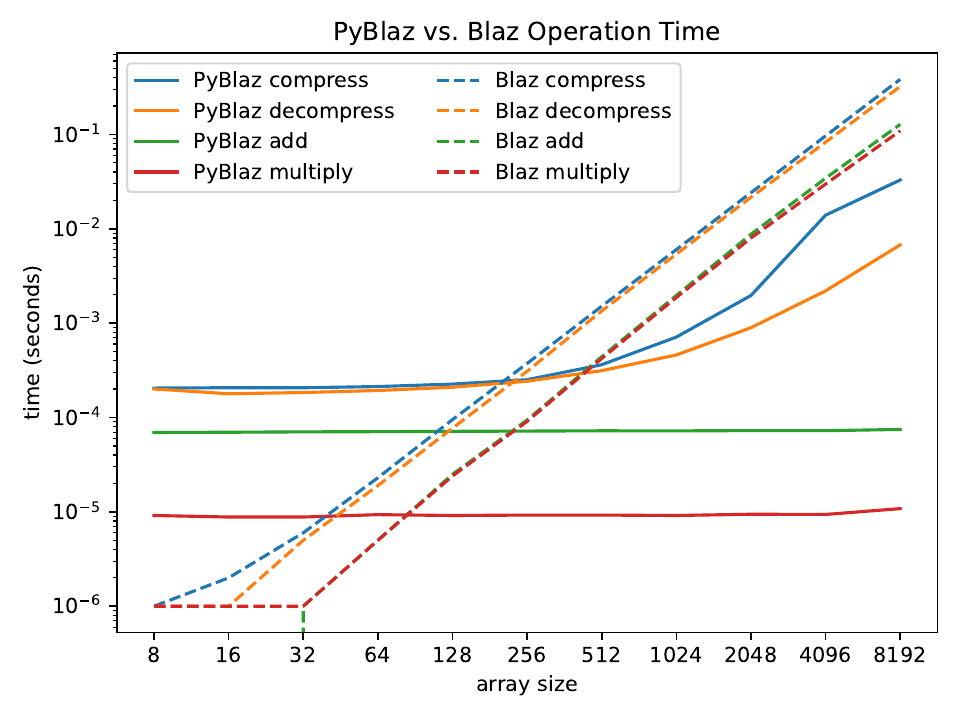} 
    \caption{
        Time taken to perform operations included in \blaz. \pyblaz compression settings were set to be comparable to those in Blaz: 2-dimensional arrays, float64 for floating-point type, int8 for index type, block shape $8 \times 8$. All arrays were square. This experiment was performed on a machine with one AMD Ryzen 5600X CPU and one NVIDIA GeForce RTX 3090 GPU.
    }
    \label{fig:timeadd}
\end{figure}

\begin{figure}
     \centering
     \begin{subfigure}[b]{0.48\textwidth}
         \centering
         \includegraphics[width=\textwidth]{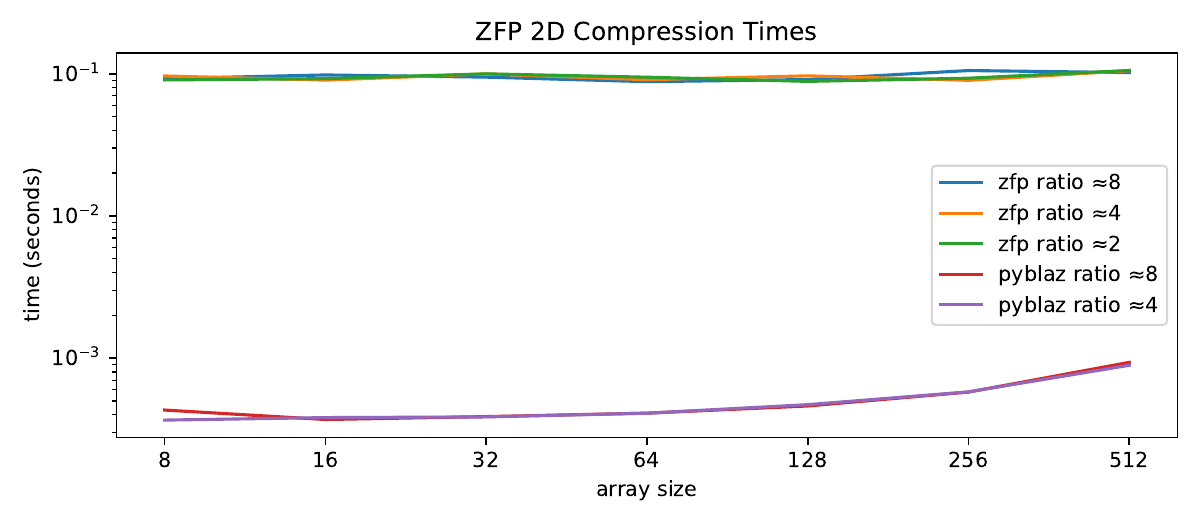}
         \subcaption{2-dimensional compression}
     \end{subfigure}
     \hfill
     \begin{subfigure}[b]{0.48\textwidth}
         \centering
         \includegraphics[width=\textwidth]{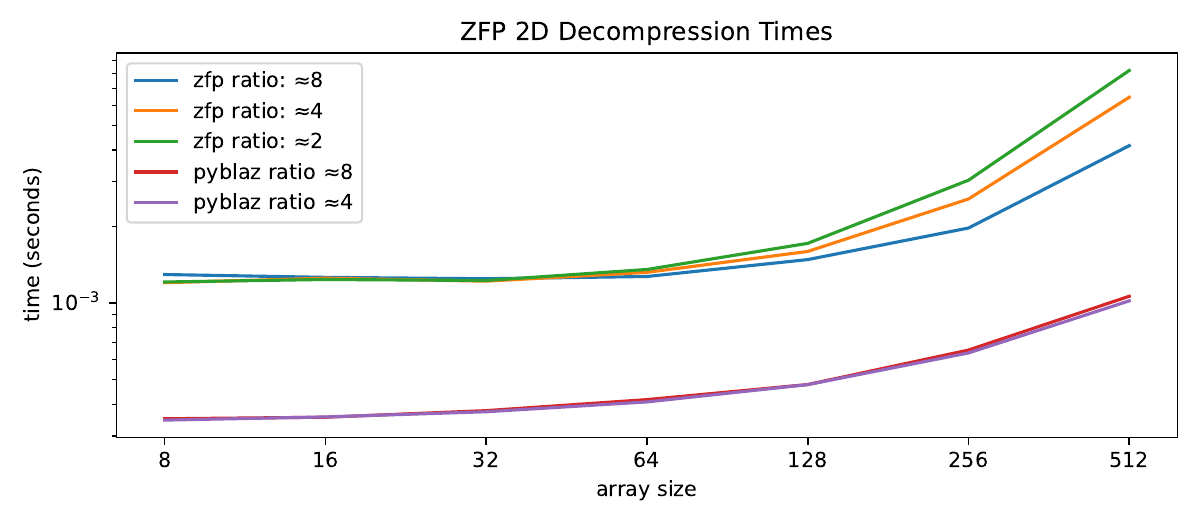}
         \caption{2-dimensional decompression}
     \end{subfigure}
     \hfill
     \begin{subfigure}[b]{0.48\textwidth}
         \centering
         \includegraphics[width=\textwidth]{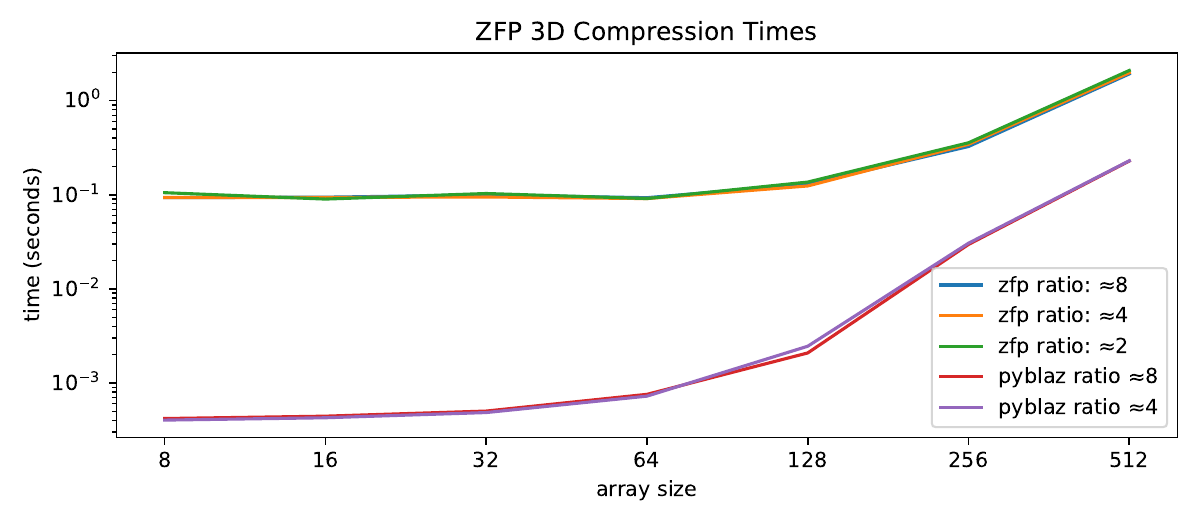}
         \caption{3-dimensional compression}
     \end{subfigure}
     \hfill
     \begin{subfigure}[b]{0.48\textwidth}
         \centering
         \includegraphics[width=\textwidth]{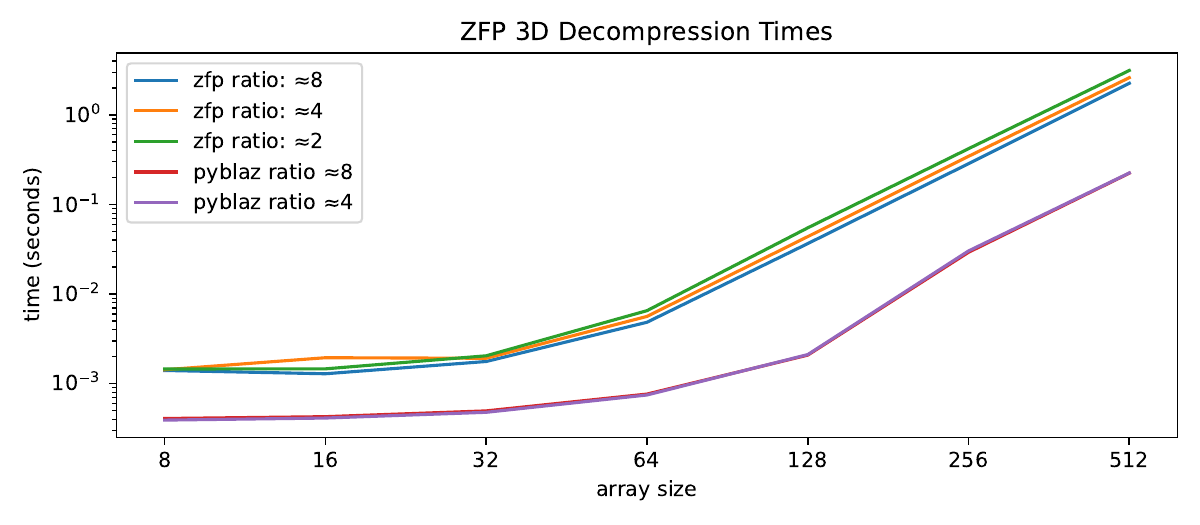}
         \caption{3-dimensional decompression}
     \end{subfigure}
     \caption{
      Compression and decompression time taken compared to ZFP using CUDA. ZFP with CUDA supports only arrays of up to 3 dimensions and compression using fixed-rate mode. ZFP decompression does not use the GPU. ZFP compression ratios of approximately 8, 4, and 2 were specified using 8, 16, and 32 bits per scalar. \pyblaz ratios of approximately 8 and 4 were achieved using bin index types \byte and \short. This experiment was performed on a machine with an AMD Ryzen 5 3600 CPU and an NVIDIA GeForce RTX 2070 Super GPU.}
     \label{fig:timevszfp}
\end{figure}

An expected benefit of \pyblaz using the GPU is higher throughput compared to the single-threaded \blaz. Figure~\ref{fig:timeadd} shows the typical behavior of compressed-space operation time in \pyblaz, with comparisons to that in \blaz.
The time taken remains constant until the GPU threads are saturated, after which it scales polynomially with array size, like the single-threaded \blaz. 
We include additional time figures for 3-dimensional arrays using a conservative block shape and various other compression settings in Appendix~\ref{sec:additionaltimefigures}.
We also compared the compression and decompression speeds of \pyblaz and those of SZ and ZFP.
We compressed and decompressed hypercubic arrays with elements ranging from 0 to 1 arranged in a constant gradient from the lowest indices to the highest indices. That is, we used the array $X$, shaped $\mathbf{s}$, such that $X_\mathbf{x} = \frac{\sum (\mathbf{x} - 1)} {\sum (\mathbf{s} - 1)}$ for all indices $\mathbf{x}$. Figure~\ref{fig:timevszfp} shows the time taken compared to ZFP.

% All time measurements shown are the average of 5 runs after 1 warmup run.

\section{Experiments and Results}
\label{sec:experimentsandresults}

% We introduce experiments that show how \pyblaz can be used on different datasets and how the compressed-space operations can be used on them.
% %
% In particular, we show the compressed-space distance and similarity measures that can be used to gain insights in the following datasets:

\noindent We study three different practical situations using \pyblaz. 
 
\vspace{1ex}
\noindent \textit{Shallow water simulation\/} is particularly used to show how errors caused due to precision tuning can be captured using our compressed-space operations. We use the simplest difference operation on each pixel of the data (done using the negation and addition compressed space operations) to show this error.
    
\vspace{1ex}
\noindent{\textit {MRI}:\/} Next, we experiment with brain MRI images in the \textit{LGG segmentation data}, showing absolute error and relative error between uncompressed and compressed mean, variance, and L$_2$ norm of one MRI image, and SSIM between two images.

\vspace{1ex}      
\noindent{\textit {Fission Data}:\/} Finally, we experiment with \textit{plutonium atom fission data} to show how the compressed-space Wasserstein distance can be used to capture topological features (e.g. a scission point). A scission point is a time interval in which the atom's nucleus splits.
 This is achieved by compressing data at each time step and then comparing data using the $L_2$ norm and Wasserstein distance in adjacent time steps. We also compare the use of these two measures.

% \vspace{1ex}
% \noindent\textit{The MAE metric:\/}
% Before one trusts direct compressed-space operations, one has to characterize compression-induced error as a function of data sources and compression settings.
%
% We provide an analysis of error resulting from compression as well as from compressed-space operations compared to uncompressed-space operations. 
%
% We employ MAE (mean absolute error) when the usefulness of the data degrades linearly with point-wise error.
% \ggcmt{Say where all MAE used later; or point to 1-2 cases.}

\subsection{Shallow water simulation data}
% \tacmt{Changed the text, please proofread}
\label{subsec:Shallowwater}
%simulations

%
\begin{figure*}[tbp]
    \centering
    \includegraphics[width=\textwidth]{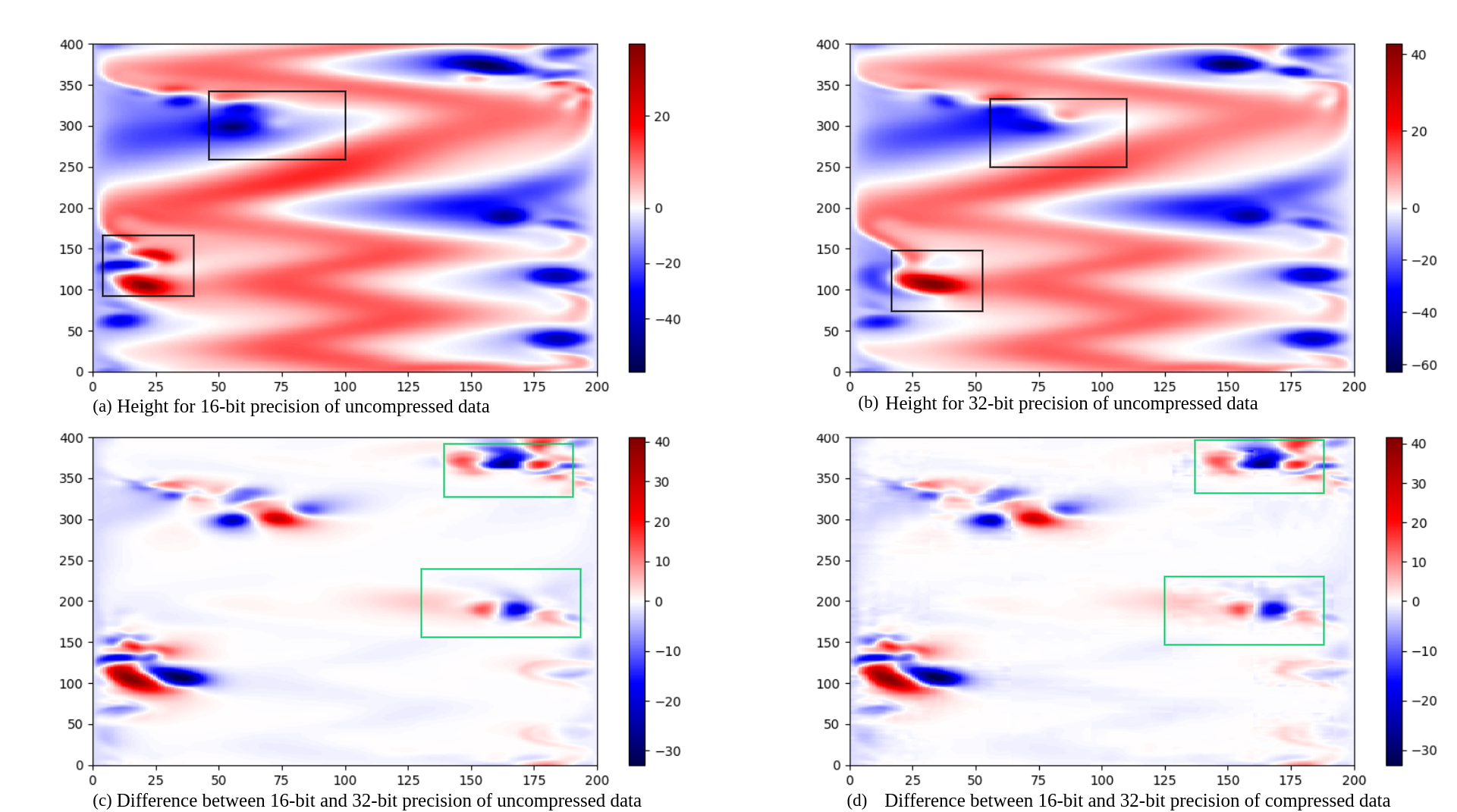}
    \caption{Height of the water surface at one-time step from a \sws using different precisions. (a) Surface height using \half and (b) Surface height using \single. These visualizations show the areas affected by the change in precision, with immediately visible differences marked with black rectangles in (a) and (b). By finding the difference between these outputs, we also capture other areas that have major differences in (c) and (d) marked with green rectangles. We also show that \pyblaz is able to capture similar differences from compressed data. Note that the surface height from this simulation could be negative.}
    \label{fig:shallowwater}
\end{figure*}
% \ggcmt{This section starts without any warning as if a gun-shot went off "boom"! Don't start a section with "Even though". Check every section. 
% You should ease into a section. 
% Say something like "Wasserstein-like measures are good for studying the impact of precision which may destroy local features etc."
% Also you are apologizing for using FP16.
% Just say "you wanted to try and see what errors we get {\em should one go to FP16} and say that your methods helped you study it using SSIM.
% }
The \swe are a set of partial differential equations derived from Navier-Stokes equations that describe the fluid flow below the pressure surface in a 2-dimensional confined boundary domain. 
These simulations can output data on multiple simulation features over time, including water velocity, depth, pressure, and height of the surface. 

We use a Julia implementation of the \sws~\cite{Shallowwater,ShallowWaters.jl} that supports different precisions including \half, \single, and \double.
%
% This package can be extended to suit specific applications. The simulation model can be implemented using a finite-volume method. The model can also work with irregular geometries by using unstructured meshes. 
%
Even though 16-bit floating-point numbers (\half) could be appropriate for specific applications, their precision is sometimes insufficient for certain scientific simulations compared to \single or \double.
Using \half could present difficulties related to the stability or accuracy of the simulations, which could lead to numerical errors or instability in subsequent calculations. 
We first verify that some imprecision occurs due to different floating-point types by looking at uncompressed simulation outputs. Then, we try to capture it by using aggressive compression settings in \pyblaz and checking whether we can achieve similar results with \pyblaz as well.
\paragraph{Dataset} 
For our experiments, we used a 500-day, double-gyre simulation with a domain shaped $200 \times 400$, with 100 grid cells in the first dimension. We varied the data type between \half and \single with boundary condition set to nonperiodic, wind forcing in x direction as double gyre and topography as seamount. More details on each of the parameters can be found in~\cite{Shallowwater}.
\paragraph{Experiment} The experiment highlights the importance of carefully choosing the precision settings in numerical simulations, and of considering the potential effects of changing precision on the \sws. We used the water surface height data obtained from the output of two simulations using \half and \single as data type. We first visualized the raw output to compare the effect of precision. 
We then visualized the element-wise difference in the data between both uncompressed forms and used compressed-space negation and addition to see the difference between the compressed forms. 
The compressor block shape is set to $16 \times 16$ for this experiment, the floating-point type is \single, and the bin index type is \byte.
\paragraph{Observation and Result} Figure~\ref{fig:shallowwater}(a) and (b) show the surface height at one simulation step using different precision settings (\half and \single), while Figure~\ref{fig:shallowwater}(c) and (d) show the difference between the two precision settings using uncompressed and compressed calculations, respectively. The black rectangles in Figure~\ref{fig:shallowwater}(a) and (b) highlight the areas with the biggest perturbations.
The perturbation caused by changing precision is not limited to the areas marked with black rectangles. The green rectangle in Figure~\ref{fig:shallowwater}(c) highlights an area where the perturbation is also present, as observed by calculating the pixel-by-pixel difference between Figure~\ref{fig:shallowwater}(a) and (b).
 With this experiment, we also show that the compressed-space difference operation (negation and element-wise addition) as shown in Figure~\ref{fig:shallowwater}(d) can be used to calculate the perturbations caused by changing precision.

The key takeaway from the experiment is that changing precision causes perturbations in both uncompressed and compressed calculations, and that compressed data and compressed-space difference operations can be used to analyze and quantify these perturbations.
This is useful because such scientific simulations have large data and are usually stored in compressed form and such lightweight compressed space operations are useful in most cases.

\subsection{LGG segmentation data}
\label{subsec:LGG}
\begin{figure}
     \centering
     \begin{subfigure}[b]{0.48\textwidth}
         \centering
         \includegraphics[width=\textwidth]{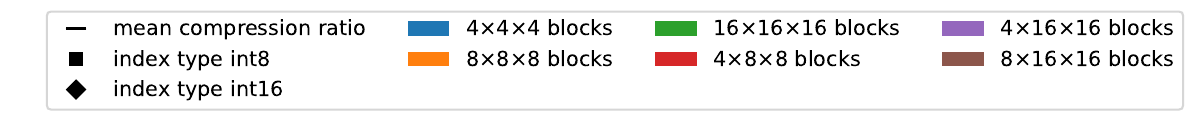}
         \subcaption{Legend}
         \label{fig:mrilegend}
     \end{subfigure}
     \hfill
     \begin{subfigure}[b]{0.48\textwidth}
         \centering
         \includegraphics[width=\textwidth]{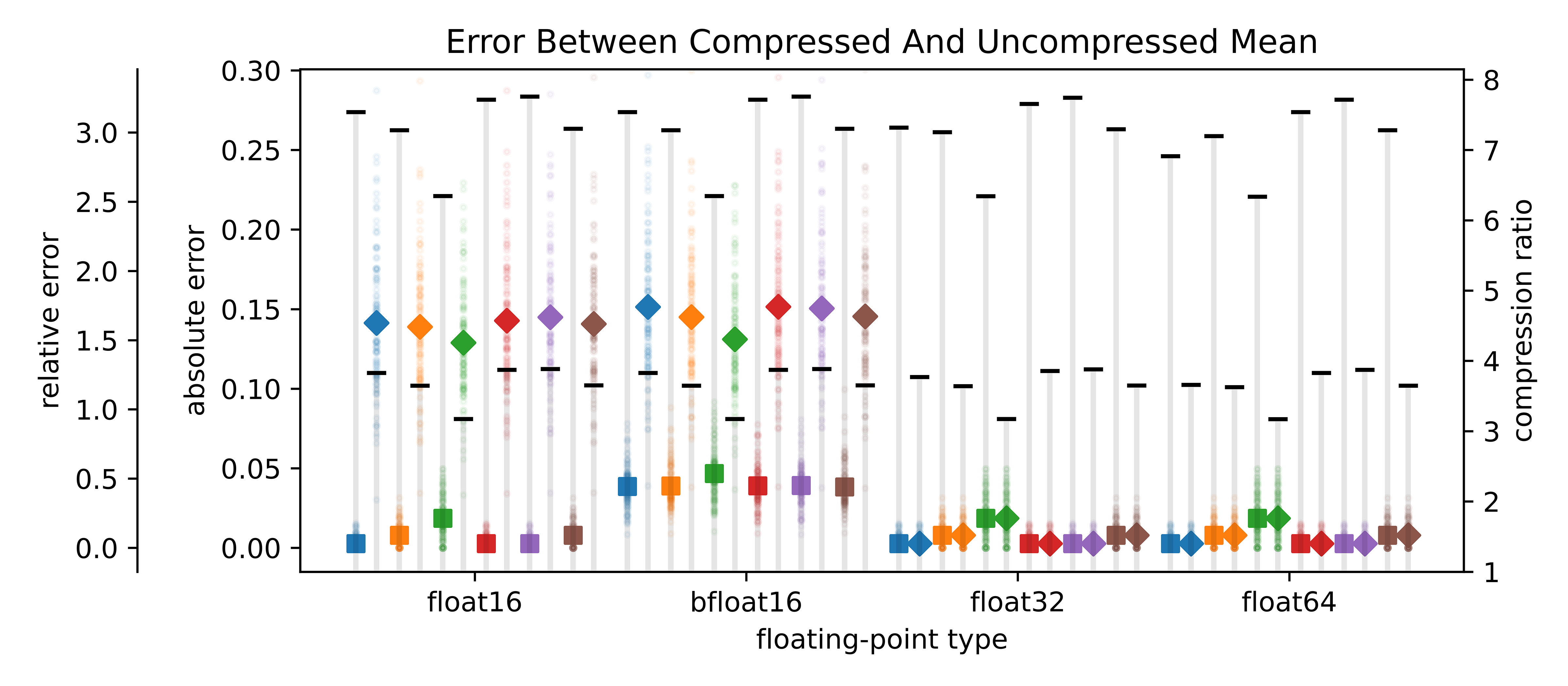}
         \caption{Mean}
         \label{fig:mrimean}
     \end{subfigure}
     \hfill
     \begin{subfigure}[b]{0.48\textwidth}
         \centering
         \includegraphics[width=\textwidth]{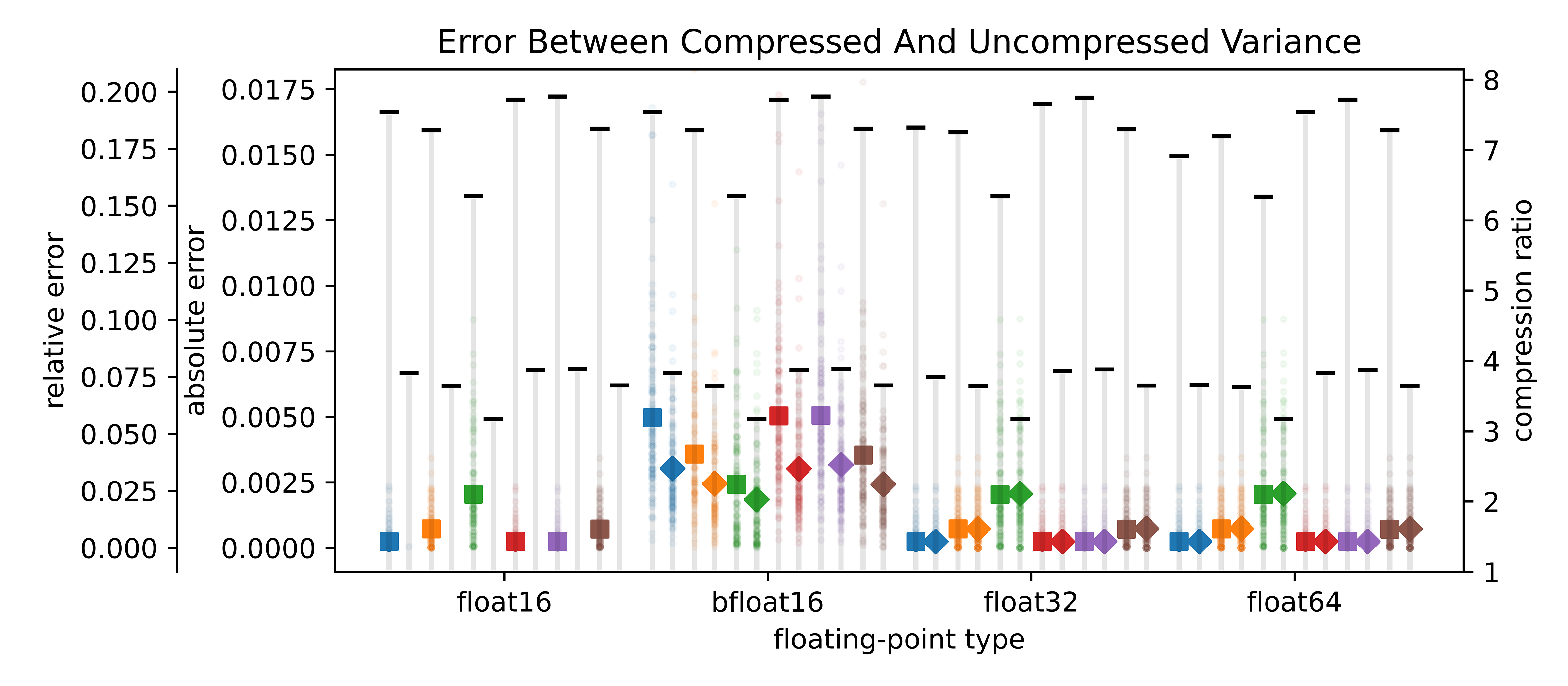}
         \caption{Variance}
         \label{fig:mrivariance}
     \end{subfigure}
     \hfill
     \begin{subfigure}[b]{0.48\textwidth}
         \centering
         \includegraphics[width=\textwidth]{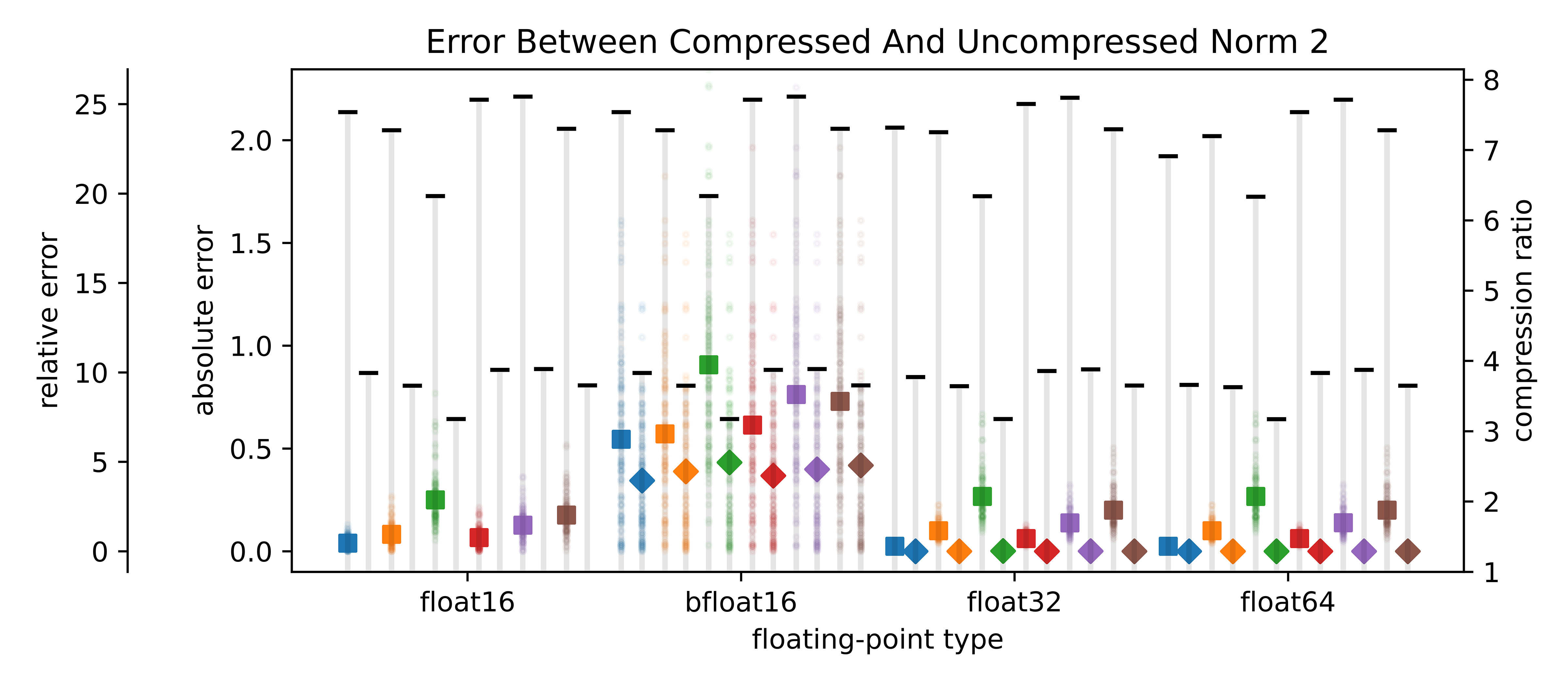}
         \caption{L$_2$ norm}
         \label{fig:mrinorm2}
     \end{subfigure}
     \hfill
     \begin{subfigure}[b]{0.48\textwidth}
         \centering
         \includegraphics[width=\textwidth]{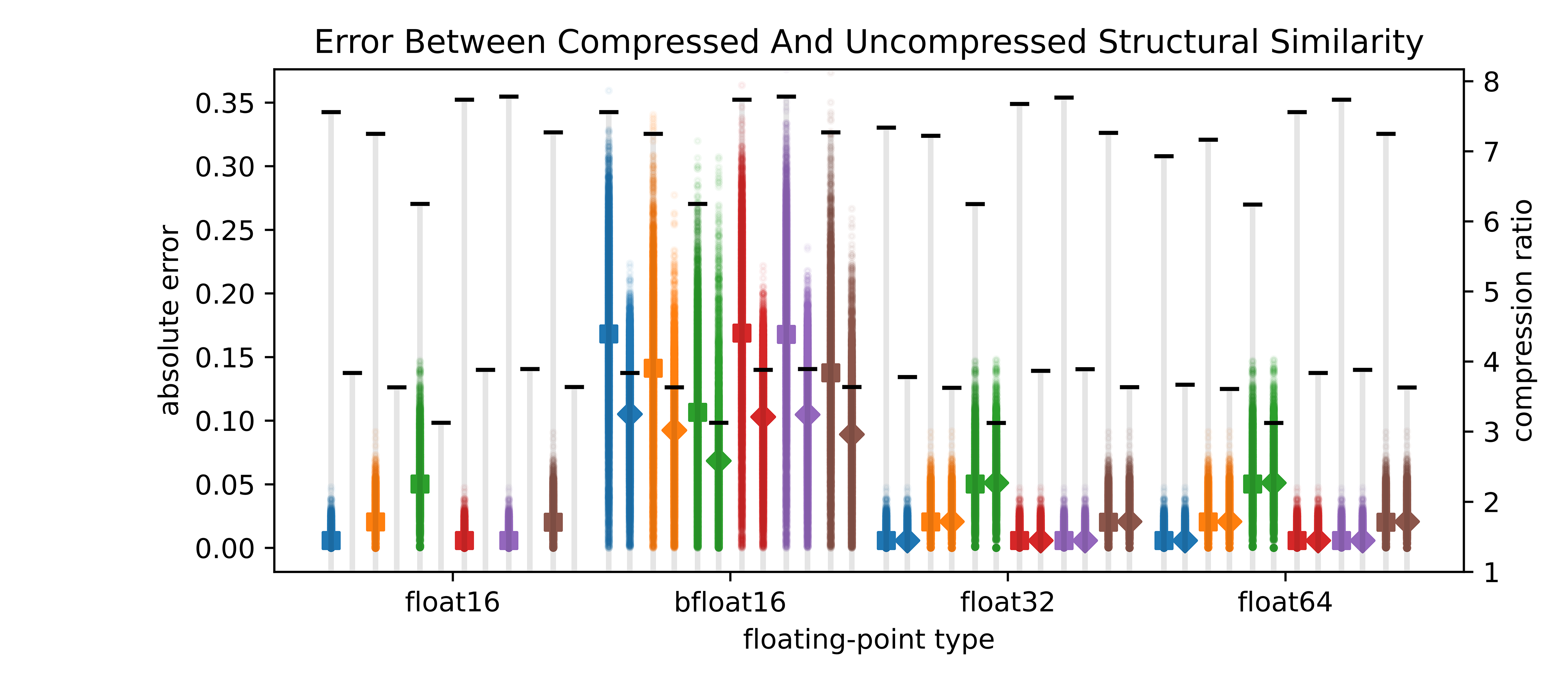}
         \caption{SSIM}
         \label{fig:mrissim}
     \end{subfigure}
     \caption{
      Absolute error and relative error between compressed-space scalar functions available in \pyblaz and uncompressed scalar functions on the FLAIR channel of the LGG segmentation dataset. Faint dots are individual examples. Squares show mean errors (MAE on the absolute axis) across all examples. Squares are missing where \nan{}s occurred on some example(s). Black horizontal lines show mean compression ratios over all examples, whose values are shown on the right vertical axis. No pruning was used. There is no relative error axis on SSIM because it is an index in [0, 1], which accounts for the magnitude of the arrays it compares.}
     \label{fig:mri}
\end{figure}
%
% \ggcmt{In the caption of Figure~4, I don't know why there is no error because "it" (SSIM?) is an index in [0,1] -- what does an index in [0,1] have to do with there being no error? Make it plainer---not needing human heads to think! This is because figures are often read independent of the body text. Good if they made sense even then.}
% \hdcmt{There is no \textbf{relative error axis} because it wouldn't make sense. Upon making this an index, the SSIM is already normalized.}
This experiment characterizes the behavior of error on four compressed-space operations as a function of different compression settings on 3-dimensional MRI images.

\paragraph{Dataset} 
The LGG segmentation dataset~\cite{mridataset} is a set of magnetic resonance imaging (MRI) images of human brains and corresponding patient data. 
Each image contains three channels: pre-contrast, fluid-attenuated inversion recovery (FLAIR), and post-contrast.
The dataset consists of 110 MRI images. The first dimension of the images, corresponding to the \textit{up} direction from the perspective of the human that was scanned, varies in size from 20 to 88, with a mean size of 35.72. The other two dimensions are constantly 256 elements in size. 
\paragraph{Experiment} We use the images in this study as an example of multidimensional arrays of spatially related values with asymmetric granularity, in that image resolution in some directions is higher than that in others, which makes these images an appropriate candidate for compression using \pyblaz. 
Because some examples had missing pre-contrast or post-contrast channels, we experimented only with the FLAIR channel.
After normalizing the values to the range [0, 1], we compressed the images in the FLAIR channel using various compression settings and measured the mean absolute and relative error between certain \pyblaz scalar functions and those on uncompressed images using plain PyTorch. The relative errors are relative to a FLAIR mean of 0.0870, averaged across each image and then across all images.
% Keep the previous sentence here. It tells them what "relative" means.
The FLAIR standard deviation is 0.1238.

The operations used in this study
were mean, variance, $L_2$ norm, and SSIM. 
We took the first three measures on individual images. We calculated the SSIM between all pairs of images in the dataset, 
% \tacmt{Can we remove the statemenr "for each image..." as it is confusing and let it be like that. Or we can frame another statement}(for each image, for each next image to the end, to the end)
cropping or padding one of them to match shapes if necessary.

\paragraph{Observations and Result} We show mean absolute errors, mean relative errors, and compression ratios in Figure~\ref{fig:mri}.
Among the floating-point types, \single and \double achieved almost the same error, while \half and \bfloat errors were probably unacceptable for many applications.
Among the 16-bit types, \half usually achieved lower error than \bfloat from its longer significand, 
% significand https://en.wikipedia.org/wiki/Significand
while \bfloat avoids \nan{}s because of its longer exponent. 
Among the \single and \double results, the lowest error is achieved using the smallest blocks and index type \short. However, depending on the function used, \byte may achieve the same error while approximately doubling the compression rate.
Also, we can see the advantage of non-hypercubic blocks. When compressing large arrays whose size is similar in all directions, we usually expect bigger block sizes to have a higher compression ratio, as fewer floating points need to be stored.
However, because the first dimension of the images in the dataset varies in size, and is usually about an eighth the size of the other dimensions, block sizes that have bigger first dimensions often take up significant additional space for padding.

Thus, for this dataset, the highest compression ratios are achieved using the non-hypercubic $4 \times 16 \times 16$ blocks, which also achieve lower error than the hypercubic $8 \times 8 \times 8$ blocks on the mean, variance, and SSIM functions.

\subsection{Plutonium atom fission data}
\label{subsec:Plutonium}
Nuclear density functional theory (DFT) is an approach used to study the properties inside an atom's nucleus.
% \hdcmt{What theory}
% \tacmt{Updated text}
Nuclear fission is a phenomenon that takes place in a nucleon-nucleon interaction in atomic nuclei. 
The process of nuclear fission involves the division of an atom's nucleus into two or more parts.
During this process, the nucleus is stretched, which causes deformation and changes in the nucleus structure. 
Accurately identifying the nuclear scission point i.e. the time steps between which the nucleus splits is a significant issue in nuclear fission research.
At the point when nuclear scission happens, the atom's topology changes, which leads to a high-magnitude change in data values.
%
% Physicists try to look for nuclear scission through a visual perspective, which is an expensive process both in time and in storage.
% And here we do that too -> no advantage
Due to the large floating-point values generated during nuclear fission, it is advantageous to compress the data and to look for the nuclear scission in compressed space.
In this experiment, we show how we can compress the data and perform a basic compressed-space operation to identify the time steps between which the scission occurred.

\begin{figure}[ht]
    \centering
    \begin{subfigure}[b]{0.48\textwidth}
        \includegraphics[width=\textwidth]{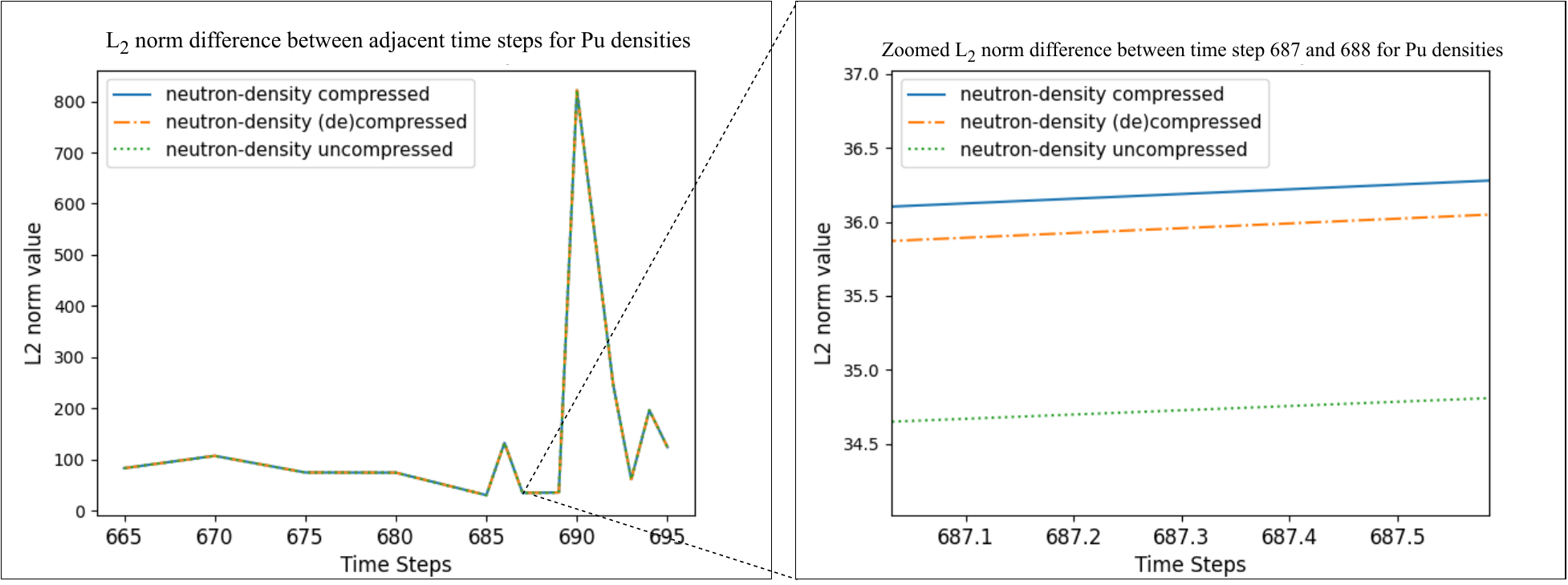}
        \caption{$L_2$ difference for uncompressed, (de)compressed, and compressed data, showing that the error between each of them is almost negligible.
        %\hdcmt{What kind of error?}
        A zoomed version of $L_2$ norm between time steps 687 and 688 is shown on the right side to show that uncompressed, (de)compressed, and compressed values are not exactly the same and actually differ by a small value. $L_2$ norm does not eliminate other peaks (noise) captured at different time steps}
    \label{fig:L2_plutonium}
    \end{subfigure}
    \hfill
    \begin{subfigure}[b]{0.48\textwidth}
    \centering
        \includegraphics[width=\textwidth]{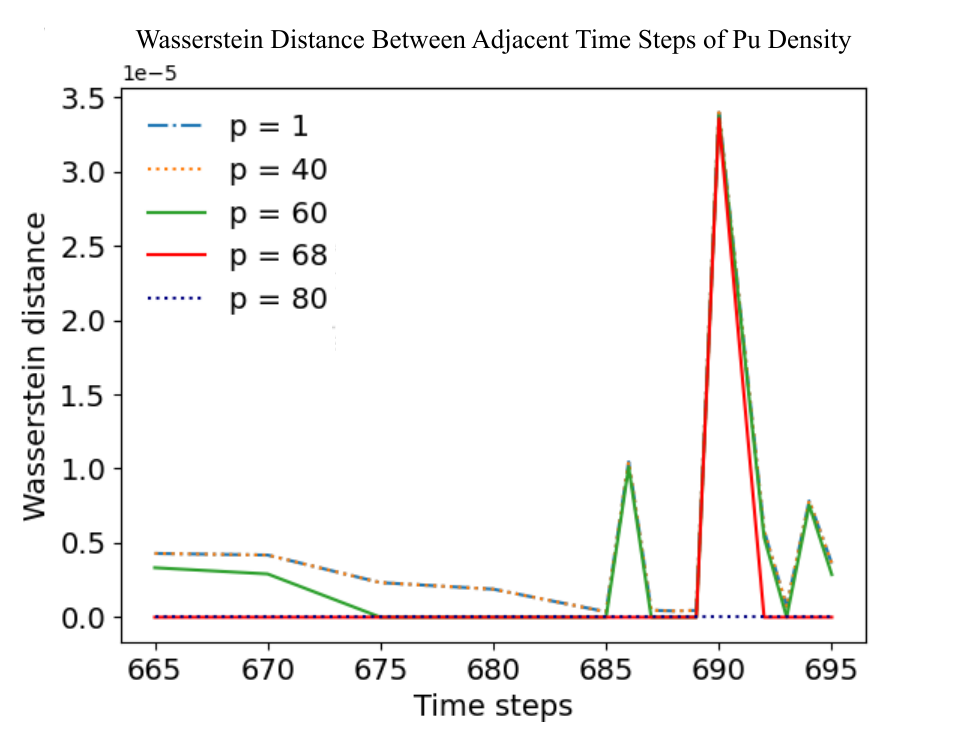}
    \caption{Approximate Wasserstein distance with different values of $p$. With $p = 68$, we clearly differentiate the nuclear scission time step from others.}
    \label{fig:wass_plutonium}
    \end{subfigure}
\caption{(a) $L_2$ norm difference between adjacent time steps for negative log-transformed plutonium neutron density for compressed and uncompressed data. We observe a major change at time step 692 showing that the nuclear scission happened at this time step. But there are other peaks that might be misleading. (b) Approximate Wasserstein distance between adjacent time steps. We see that the misleading peak starts to vanish as the order ($p$) of the Wasserstein distance changes and only one peak is left when $p = 68$ (shown with the red line), showing only major changes (or scission point) in the data. A red line graph with $p = 68$ is also shown, clearly showing one major peak captured using the Wasserstein distance.}
\label{fig:plutoniumresults}
\end{figure}

\paragraph{Dataset} The dataset consists of spatial densities in plutonium atoms that include the spatial densities of neutrons in the nucleus. These densities are sampled on a grid of shape $40\times40\times66$.
The dataset is negative log-transformed\footnote{Negative log-transformation is a method of normalizing and managing negative values in a dataset. A constant is added to all the data values, and then log transformation is performed on them. Since this process maintains the relative order of the values and an analogue of their relative magnitudes, it does not affect the overall behavior of the data and hence a similar effect would appear in the data if the experiments were performed on raw densities.} and sampled at 15 different time steps, specifically [665, 670, 675, 680, 685, 686, 687, 688, 689, 690, 692, 693, 694, 695, 699]. 
Previous work shows that the nuclear scission happens between time steps 690 and 692~\cite{JCN, topology, MRRS}.
\paragraph{Experiment} We take the negative log-transformed neutron densities of every time step, each provided as a 3-dimensional array, and compress each array individually. To find the nuclear scission point we find the $L_2$ norm of the difference in log density between adjacent time steps $D_1$ and $D_2$ as
$\lVert D_1 - D_2 \rVert_2$.
Figure~\ref{fig:L2_plutonium} shows these differences per time step pair.
Since the $L_2$ norm was not able to capture the nuclear scission point with certainty (as we explain later),
we also calculate the approximate Wasserstein distance between each time step and plot the line graph with different values of the order ($p$).
Although the higher-order norms such as $L_\infty$ are also able to ignore the noise. We used the approximate Wasserstein distance to demonstrate a use for approximations in \pyblaz.

For our experiment, we used a block size of $16 \times 16 \times 16$, index type \short, and \single as the data type for the compressor.
%\hdcmt{What was the index type?}
%

%
% \tacmt{Rewritten the observation and results, please proof read}
\paragraph{Observation and Results} The $L_2$ norm differences and Wasserstein distances between adjacent time steps are shown in Figure~\ref{fig:plutoniumresults}. A major peak is obtained between time steps 690 and 692 when operating on compressed data---{\em matching previously known results.}

We first observe the results obtained using $L_2$ norm with block size $16 \times 16 \times 16$. Although a major peak is shown at time step 690 (indicating a scission event between time steps 690 and 692), we also observed other nearby peaks between time steps 685 and 686 and between steps 695 and 699. Existing literature shows that other peaks are noise, and the topology of the data is not changing at those points ~\cite{JCN, topology}. These noise peaks make the scission less certain.
% We also experimented with the $L_2$ norm between adjacent time steps by changing the block sizes and data types in the compression settings.
Changing compression settings did not change this behavior.
% We obtained similar results and hence it is hard to differentiate whether these peaks are noise or there is a change in topology.
The compressed-space $L_2$ norm maintains sufficient fidelity to capture topological changes of interest because the \textit{maximum} error between the $L_2$ norms of each time step from uncompressed and compressed data is approximately 1.68 (as shown in the right side of Figure ~\ref{fig:L2_plutonium}), while the mean $L_2$ norm is 618.97.

Using the $p$-order Wasserstein distance between adjacent time steps (Figure~\ref{fig:wass_plutonium}, we observe that as $p$ increases, the smaller peaks are suppressed, and only one major peak (where nuclear scission occurred) is first observed when the order is set to 68. 
We also observe that if the order $\ge$80 all the peaks vanish. This shows that (approximate) compressed-space operations are able to identify nuclear scission and that one distance measure has advantages over another in certain cases.
%

% We verify our results by calculating the $L_2$ norm on uncompressed data. We obtain a similar trend in the difference metric, showing that the results obtained in the compressed domain are reliable.

\section{Conclusions}
\label{sec:conclusion}

In this paper, 
We studied many direct operations on compressed data,
selecting the appropriate distance metrics to  characterize errors on top of compression-decompression errors (if any).
Using a \sws,
we showed how operations such as negation and element-wise addition can be used on large floating-point data of different precisions.
  We also show how large block sizes can be used while still capturing
  precision-change induced  perturbations in the dataset.
We then characterized the error between compressed-space scalar functions and their uncompressed-space counterparts as a function of data from the LGG segmentation dataset and compression settings.
%
% In particular we showed a case  involving non-hypercubic block shapes.
%
We finally
used the physics dataset of nuclear fission of plutonium atoms to show the change in the topology of time-series data, 
illustrating
the merits of using of
the Wasserstein distance over the $L_2$ norm. 
This validates the promise of \pyblaz, the first data compressor supporting a dozen compressed domain operations.

\paragraph*{Future work}
We did not have an occasion to illustrate most of the supported operations through experiments, leaving them for future evaluation.
Incorporating compression or decompression
 might be controlled by
  \texttt{pragma}s placed against their data structure declarations.
Another usage scenario is in ensemble-testing where
applications are compiled under different flags
for testing the numerical characteristics of the code, as in \cite{ahn2021keeping}.
In \cite{ahn2021keeping}, only simplistic distance measures were used and the distance calculation was in the uncompressed domain. 
With \pyblaz, we can employ more sophisticated measures while keeping the time-sequences of evolving simulation results in compressed form.
\pyblaz can be made to automatically change its compression settings in order to enforce some $L_\infty$ error bound through Bayesian optimization or a similar search process instead of relying on the user to find optimal compression settings.
We plan to develop
rigorous stage-wise error analysis
 for \pyblaz similar to what has been done for ZFP \cite{fox-zfp-error}.

Last but not least, formal verification of compression, decompression, and compressed-space operations is almost a requirement, because subtle flaws might look confusingly similar to actual data aberrations.
This verification would be necessary both at the higher-level (e.g., by coming up with equational axioms pertaining to various operations).
Code-level verification is equally important: an off-by-one error again might not cause a visible alarm until when one inadvertently handles the wrong (and critical) data.

\vspace{1ex}

\noindent{\bf Acknowledgements:}
Supported in part by 
 NSF Awards 2124100, 1956106, 2124100 and 2319507,
and also supported by the U.S. Department of Energy, Office of Science, Office of Advanced Scientific Computing Research under
 award DE-SC0022252.

\bibliographystyle{IEEEtran}
\bibliography{references,compression-and-fp-bibs}

\section*{Appendix}

\subsection{Example orthonormal transform}
\label{sec:dctexample}
Suppose we wish to perform DCT on a blocked array $B$ using block shape (4, 8). Recall that $\{\mathbf{H}_1, \dots, \mathbf{H}_{d} \}$ are the DCT matrices for each block size and
$\mathbf{H}_{ij} = \sqrt{\frac{1 + (j>1)}{s}} \cos \frac{ \pi i (2j+1)}{2s}$. In this case, we have two DCT matrices $\mathbf{H_1}$ and $\mathbf{H_2}$, one for each dimension, such that
\mbox{$\mathbf{H_1} = \begin{bmatrix}
\sqrt{\frac{1}{4}} \cos \frac{0 \pi}{8} & \sqrt{\frac{2}{4}} \cos \frac{0 \pi}{8} & \sqrt{\frac{2}{4}} \cos \frac{0 \pi}{8} & \sqrt{\frac{2}{4}} \cos \frac{0 \pi}{8}\\ 
\sqrt{\frac{1}{4}} \cos \frac{1 \pi}{8} & \sqrt{\frac{2}{4}} \cos \frac{3 \pi}{8} & \sqrt{\frac{2}{4}} \cos \frac{5 \pi}{8} & \sqrt{\frac{2}{4}} \cos \frac{7 \pi}{8}\\ 
\sqrt{\frac{1}{4}} \cos \frac{2 \pi}{8} & \sqrt{\frac{2}{4}} \cos \frac{6 \pi}{8} & \sqrt{\frac{2}{4}} \cos \frac{10 \pi}{8} & \sqrt{\frac{2}{4}} \cos \frac{14 \pi}{8}\\ 
\sqrt{\frac{1}{4}} \cos \frac{3 \pi}{8} & \sqrt{\frac{2}{4}} \cos \frac{9 \pi}{8} & \sqrt{\frac{2}{4}} \cos \frac{15 \pi}{8} & \sqrt{\frac{2}{4}} \cos \frac{21 \pi}{8}\end{bmatrix}$}
and \\
\mbox{$\mathbf{H_2} = \begin{bmatrix}
\sqrt{\frac{1}{8}} \cos \frac{0 \pi}{16} & \sqrt{\frac{2}{8}} \cos \frac{0 \pi}{16} & \dots & \sqrt{\frac{2}{8}} \cos \frac{0 \pi}{16}\\ 
\sqrt{\frac{1}{8}} \cos \frac{1 \pi}{16} & \sqrt{\frac{2}{8}} \cos \frac{3 \pi}{16} & \dots & \sqrt{\frac{2}{8}} \cos \frac{15 \pi}{16}\\ 
\vdots & \vdots & \ddots & \vdots\\
\sqrt{\frac{1}{8}} \cos \frac{7 \pi}{16} & \sqrt{\frac{2}{8}} \cos \frac{21 \pi}{16} & \dots & \sqrt{\frac{2}{8}} \cos \frac{105 \pi}{16}
\end{bmatrix}$.}

In $d=2$ dimensions, we need the symbol space $\Sigma$ to have at least $2d=4$ symbols to use as indices. The symbols $\{\alpha, \beta, \gamma, \delta\}$ will do.

If $B$ is the blocked array, and $B_\mathbf{k}$ is some block $\mathbf{k}$, we can express a block of transformed coefficients as $C_\mathbf{k}$, whose each element indexed $( \gamma \delta)$ is
$C_{\mathbf{k} \gamma \delta} =
\sum_{\alpha = 1}^{4} 
\sum_{\beta = 1}^{8} 
B_{\mathbf{k} \alpha \beta} 
\mathbf{H}_{1 \; \alpha \gamma} 
\mathbf{H}_{2 \; \beta \delta}: \gamma \in [1..4], \delta \in [1..8]$. Using Einstein's summation notation, the summation symbols and index ranges are omitted, so we have
\mbox{$C_{\mathbf{k} \gamma \delta} =
B_{\mathbf{k} \alpha \beta} 
\mathbf{H}_{1 \; \alpha \gamma} 
\mathbf{H}_{2 \; \beta \delta}$}.

Then for all blocks, let $\mathbf{b}$ be the arrangement of blocks in the blocked array, so that all blocks of coefficients are \mbox{$C_{[\mathbf{1}..\mathbf{b}] \gamma \delta} =
B_{[\mathbf{1}..\mathbf{b}] \alpha \beta} 
\mathbf{H}_{1 \; \alpha \gamma} 
\mathbf{H}_{2 \; \beta \delta}$}.

Similarly in 3 dimensions, using some $\mathbf{H_3}$, $\Sigma = \{\alpha, \beta, \gamma, \delta, \epsilon, \zeta\}$, we would have \mbox{$C_{[\mathbf{1}..\mathbf{b}] \delta \epsilon \zeta} =
B_{[\mathbf{1}..\mathbf{b}] \alpha \beta \gamma} 
\mathbf{H}_{1 \; \alpha \delta} 
\mathbf{H}_{2 \; \beta \epsilon}
\mathbf{H}_{3 \; \gamma \zeta}$}.

Then, generalizing to arbitrary dimensions, we have
$C_{[\mathbf{1}..\mathbf{b}] [\Sigma_{d+1}..\Sigma_{2d}]} =
B_{[\mathbf{1}..\mathbf{b}] [\Sigma_{1}..\Sigma_{d}]} 
\mathbf{H}_{1 \; \Sigma_{1} \Sigma_{1+d}} 
\mathbf{H}_{2 \; \Sigma_{2} \Sigma_{2+d}}
\dots
\mathbf{H}_{d \; \Sigma_{d} \Sigma_{2d}}$.

\subsection{Additional time figures}
\label{sec:additionaltimefigures}

\begin{figure*}[h]
    \centering
    \begin{subfigure}[b]{0.31\textwidth}
        \centering
        \includegraphics[width=\textwidth]{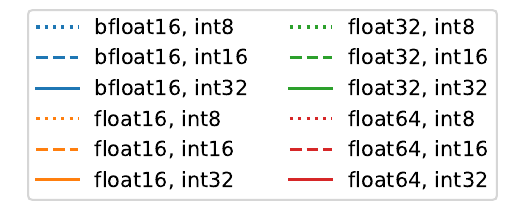}
        \subcaption{Legend}
    \end{subfigure}
    \begin{subfigure}[b]{0.31\textwidth}
        \centering
        \includegraphics[width=\textwidth]{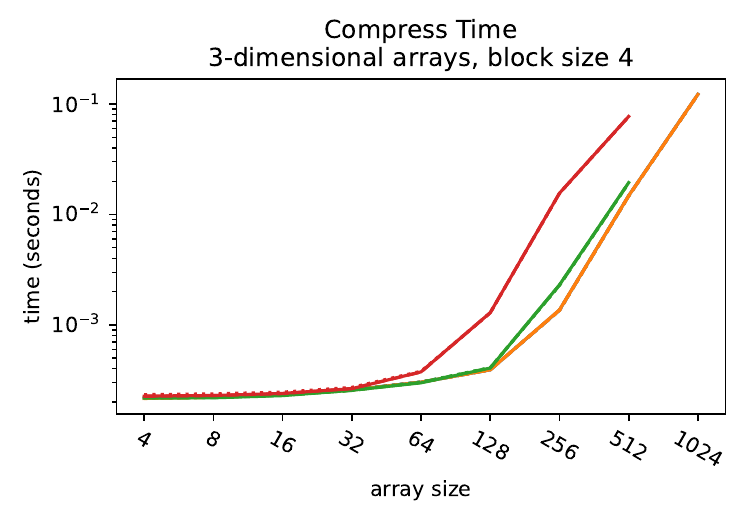}
        \caption{Compress}
    \end{subfigure}
    \begin{subfigure}[b]{0.31\textwidth}
        \centering
        \includegraphics[width=\textwidth]{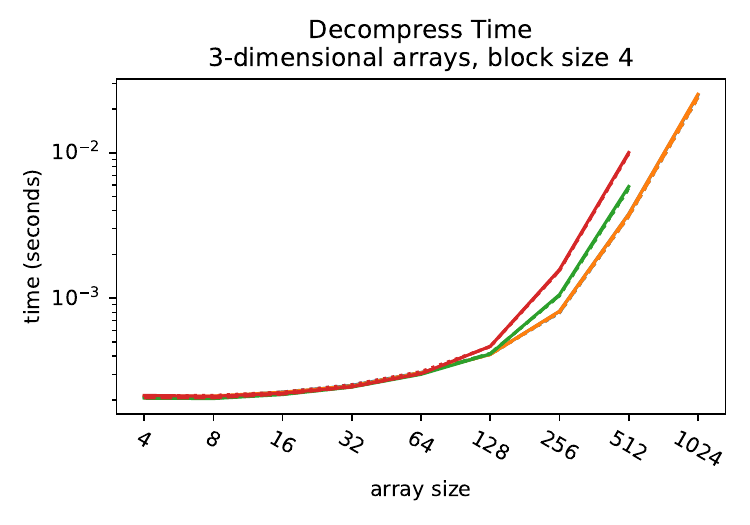}
        \caption{Decompress}
    \end{subfigure}
    \begin{subfigure}[b]{0.31\textwidth}
        \centering
        \includegraphics[width=\textwidth]{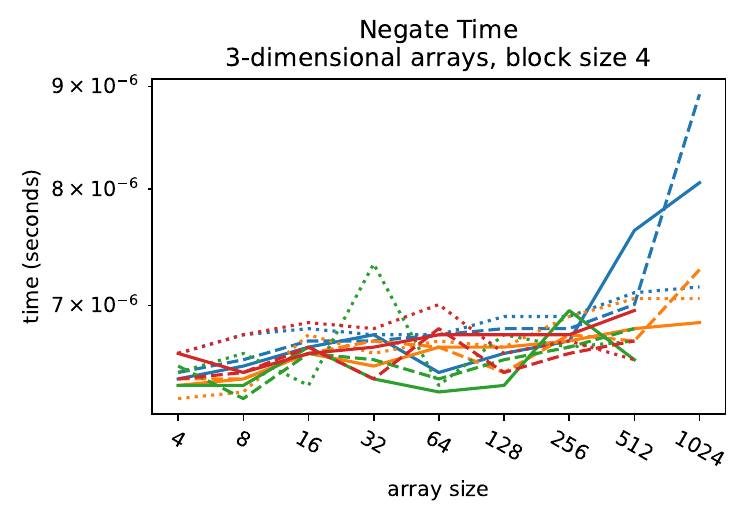}
        \caption{Negate}
    \end{subfigure}
    \begin{subfigure}[b]{0.31\textwidth}
        \centering
        \includegraphics[width=\textwidth]{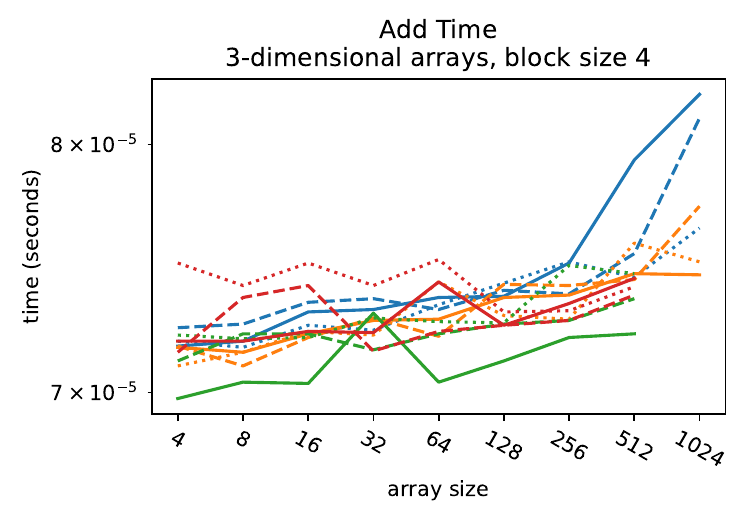}
        \caption{Add}
    \end{subfigure}
    \begin{subfigure}[b]{0.31\textwidth}
        \centering
        \includegraphics[width=\textwidth]{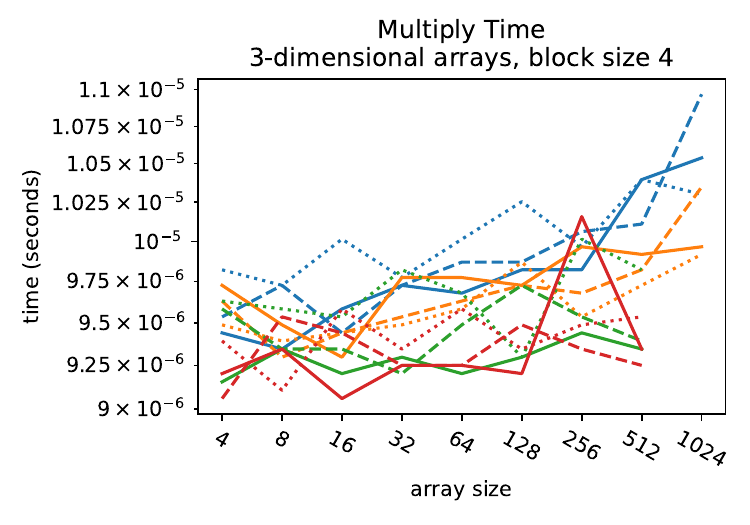}
        \subcaption{Multiply}
    \end{subfigure}
    \begin{subfigure}[b]{0.31\textwidth}
        \centering
        \includegraphics[width=\textwidth]{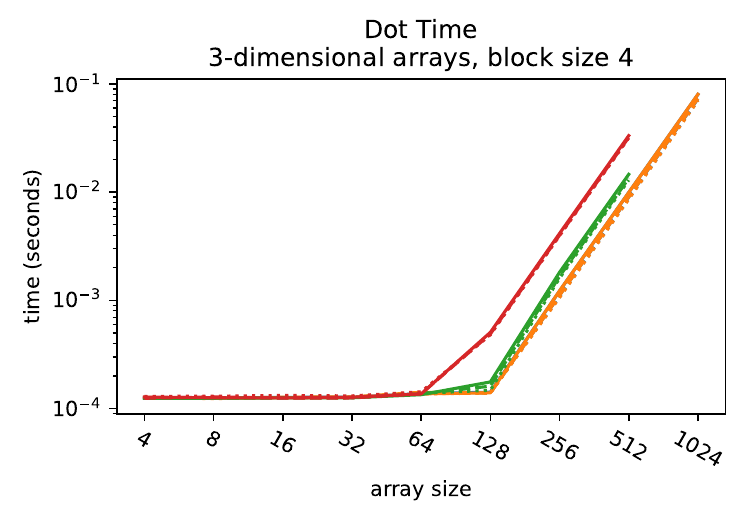}
        \subcaption{Dot}
    \end{subfigure}
    \begin{subfigure}[b]{0.31\textwidth}
        \centering
        \includegraphics[width=\textwidth]{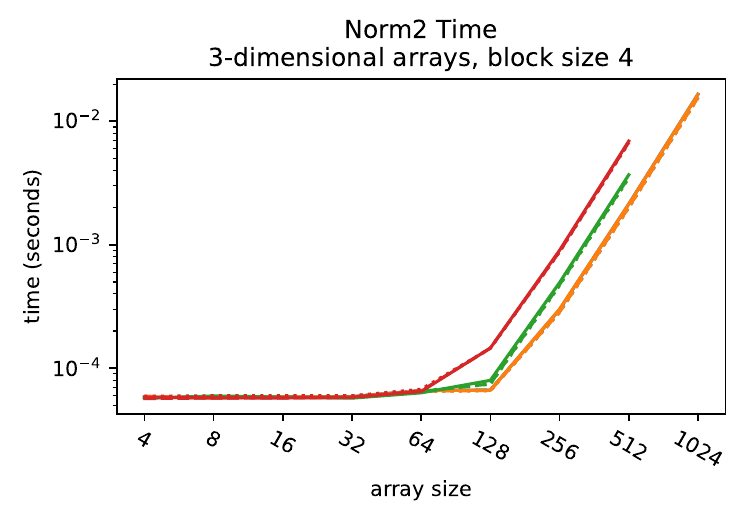}
        \subcaption{L$_2$ norm}
    \end{subfigure}
    \begin{subfigure}[b]{0.31\textwidth}
        \centering
        \includegraphics[width=\textwidth]{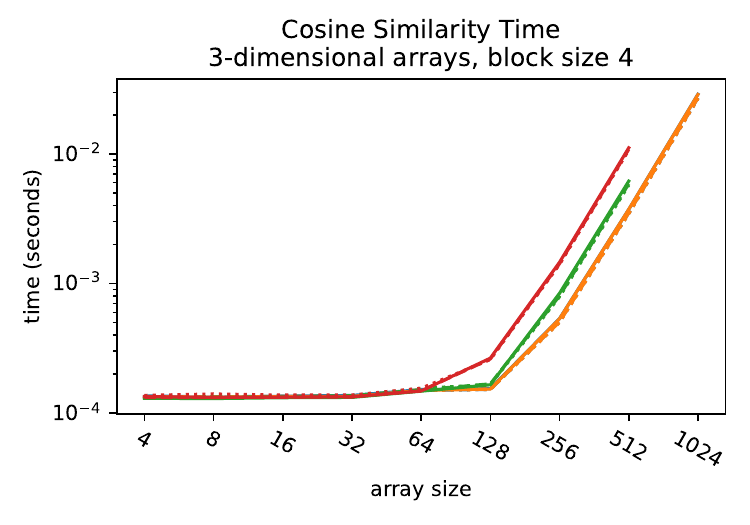}
        \subcaption{Cosine similarity}
    \end{subfigure}
    \begin{subfigure}[b]{0.31\textwidth}
        \centering
        \includegraphics[width=\textwidth]{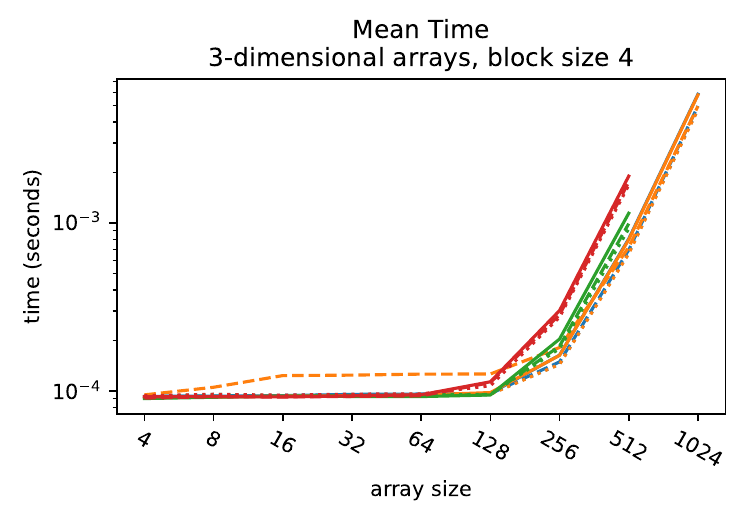}
        \subcaption{Mean}
    \end{subfigure}
    \begin{subfigure}[b]{0.31\textwidth}
        \centering
        \includegraphics[width=\textwidth]{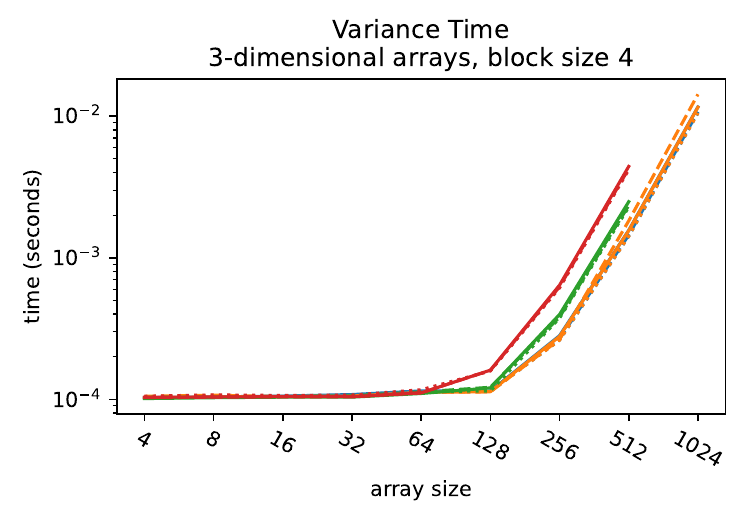}
        \subcaption{Variance}
    \end{subfigure}
    \begin{subfigure}[b]{0.31\textwidth}
        \centering
        \includegraphics[width=\textwidth]{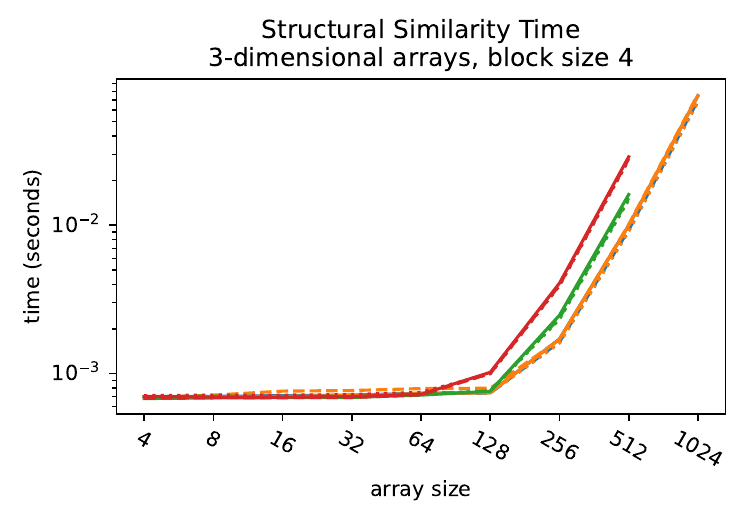}
        \subcaption{SSIM}
    \end{subfigure}
    \caption{\pyblaz operation time for cubic arrays, block size 4.}
    \label{fig:time3dbs4}
\end{figure*}

Figure \ref{fig:time3dbs4} shows the time taken to perform various compressed-space operations available in \pyblaz.

\end{document}